\documentclass[11pt]{article}
\usepackage{amsmath}
\DeclareMathOperator{\Tr}{Tr}
\usepackage{amssymb}
\usepackage{cite}
\usepackage{slashed}
\usepackage{setspace}

\usepackage{amsfonts}

\makeatletter
\@addtoreset{equation}{section}
\makeatother

\newcommand{\half}{{\textstyle{\frac{1}{2}}}}

\def\ndelta{\delta\hspace{-0.50em}\slash\hspace{-0.05em} }

\begin{document}

\begin{flushright}
\hfill{\hfill{}}

\end{flushright}

\vspace{15pt}
\begin{center}
{\Large {\bf BMS charges in polyhomogeneous spacetimes}}

\vspace{30pt}
{\bf Mahdi Godazgar and George Long}

\vspace{20pt}

{\it School of Mathematical Sciences, Queen Mary University of London, \\
Mile End Road, E1 4NS, United Kingdom.}

\vspace{35pt}
 
August 3, 2020

\vspace{160pt}

\underline{ABSTRACT}
\end{center}

\noindent 
We classify the asymptotic charges of a class of polyhomogeneous asymptotically-flat spacetimes with finite shear, generalising recent results on smooth asymptotically-flat spacetimes. Polyhomogenous spacetimes are a formally consistent class of spacetimes that do not satisfy the well-known peeling property.  As such, they constitute a more physical class of asymptotically-flat spacetimes compared to the smooth class.  In particular, we establish that the generalised conserved non-linear Newman-Penrose charges that are known to exist for such spacetimes are a subset of asymptotic BMS charges. 

\thispagestyle{empty}

\vfill
E-mails: m.godazgar@qmul.ac.uk, g.long@qmul.ac.uk

\pagebreak

\section{Introduction}
One of the most striking results in the mathematical study of gravitational waves in general relativity is the so-called peeling property \cite{bondi1962gravitational, sachs1962gravitational,newman1962approach} (see, e.g., also Ref.\ \cite{Wald}).  The peeling property is a statement regarding the asymptotic behaviour of the Weyl tensor components as one approaches null infinity.  For a smooth asymptotically-flat spacetime, the result follows from the assumed smoothness of the unphysical spacetime upon conformal compactification \cite{Penrose:1965am}.  In Bondi coordinates \cite{bondi1962gravitational, sachs1962gravitational}, it is 
\begin{equation}
 C_{abcd} = r^{-1} C^{(N)}_{abcd} + r^{-2} C^{(III)}_{abcd} + r^{-3} C^{(II)}_{abcd} + r^{-4} C^{(I)}_{abcd} +{\cal O}(r^{-5}),
\end{equation}
where $r$ is an affine parameter along an outgoing null geodesic.  The superscripts on the Weyl tensors on the RHS denote the components of the Weyl tensor in a null basis that is used to define the Petrov type of the spacetime.  Thus, the leading order term corresponds to the Weyl tensor components of Petrov type N.  Given that the Weyl tensor encompasses the remaining degrees of freedom in the curvature, the peeling property can be viewed as a statement regarding the fall-off behaviour of isolated gravitating bodies and the radiation they emit.  However, already in Ref.\ \cite{newman1962approach} (see footnote 27), doubts were expressed regarding the validity of the assumptions that lead to this result.  Since then, the question of the validity of the peeling property has attracted much research; see e.g.\ Refs.\ \cite{damour, Christodoulou, CK, bicak1993relativistic, andersson1993hyperboloidal, andersson1994hyperboloidal, chrusciel1995gravitational, ashtekar1997behavior, Friedrich:2017cjg, angelopoulos2018late, angelopoulos2019logarithmic}.  What is clear by now is that the smoothness assumption precludes many interesting physically relevant cases.  One class of spacetimes that move away from the smoothness assumption are those that admit a \emph{polyhomogeneous} expansion \cite{chrusciel1995gravitational}.  These spacetimes are formally consistent with the Einstein equations and admit the BMS group as an asymptotic symmetry group \cite{chrusciel1995gravitational}, have a well-defined Trautman-Bondi mass parameter \cite{Chrusciel:1998he} and admit Newman-Penrose charges \cite{kroon1998conserved,valiente1999logarithmic}.  Importantly, they provide an example of a more realistic class of asymptotically-flat spacetimes than the smooth case.

In this paper, we shall study the asymptotic BMS charges admitted by polyhomogeneous spacetimes that have a finite shear \cite{kroon1998conserved}.  This subset of polyhomogeneous spacetimes have a slightly better fall-off property at leading order compared with the most general spacetimes.  We will concentrate on this large subset of polyhomogeneous spacetimes in order to make the rather involved calculations tractable.  However, we are confident that the results obtained in this paper may be generalised to the full class.

In recent work, a relation has been established \cite{godazgar2019subleading, godazgar2019tower}, in the smooth case, between Newman-Penrose charges \cite{newman1968new}, which are a set of ten conserved non-linear charges at null infinity, and asymptotic BMS charges \cite{iyer1994some, barnich2002covariant, barnich2011bms}, which are the charges associated with the generators of the BMS algebra via the Noether theorem.  Although, such a relation ought to be natural, remarkably, such a relation had not been previously found.  Indeed, in order to make progress, it has been required to extend the notion of asymptotic BMS charges to include subleading charges \cite{godazgar2019subleading} and new dual charges \cite{godazgar2019new, godazgar2019tower}, which have recently been derived from first principles \cite{Godazgar:2020gqd,Godazgar:2020kqd}.  

Our aim in this work is to extend the formalism developed in Refs.\ \cite{godazgar2019subleading, godazgar2019tower} to classify the asymptotic BMS charges within the class of polyhomogeneous spacetimes with finite shear.  This generalisation is non-trivial for two reasons: the calculational complexity increases substantially when considering polyhomogenous spacetimes and particular features of the polyhomogeneous expansion raises interesting new questions about the nature of charges, as will become apparent below.  In particular, the nature of the characteristic value problem applied to polyhomogeneous spacetimes means that non-trivial conserved BMS charges can be defined in terms of initial data that do not evolve.  This is a new feature that is specific to polyhomogeneous spacetimes and compels us to reappraise what we mean by conserved BMS charges.  

Our main result is to establish a relation between asymptotic BMS charges and the generalised Newman-Penrose charges discovered in Ref.\ \cite{kroon1998conserved} for polyhomogenous spacetimes with finite shear.  Generalised Newman-Penrose charges exist for the full class of polyhomogeneous spacetimes \cite{valiente1999logarithmic} and we expect this relation to also hold in the full class.

The insights gained from this study have led to a better understanding of how Newman-Penrose charges come about and the possibility of identifying conserved charges at lower orders. An as of yet open question is whether conserved charges could exist at lower orders in the $1/r$-expansion.  We hope to tackle this interesting problem in a future work.

In Section \ref{preliminaries}, we give some prerequisite information regarding polyhomogeneous spacetimes and the fall-off behaviour of their Weyl tensor components, the Einstein equations and the action of the BMS group on the metric components.  Also, we define the subclass of polyhomogenous spacetimes with finite shear.  In Section \ref{BMS}, we classify the standard BMS charges up to order $1/r^3$ and identify a subset of five conserved non-linear charges.  Similarly, in Section \ref{Dual}, we classify the dual charges defined in Ref.\ \cite{godazgar2019tower}  up to order $1/r^3$ and, again, discover a subset of five conserved non-linear charges. In Section \ref{NP}, we show, via a translation to the Newman-Penrose formalism, that the set of ten conserved charges found in Sections \ref{BMS} and \ref{Dual} are equivalent to the generalised Newman-Penrose charges of Ref.\ \cite{kroon1998conserved}.

\section{Preliminaries} \label{preliminaries}
A polyhomogeneous spacetime is one for which the metric components can be expanded asymptotically as a combination of powers of $r^{-1}$ and positive powers of $z\equiv \log r$ as $r\rightarrow \infty$. For example, a function $f$ admits a polyhomogeneous expansion if 
\begin{equation}
f(r) = f_0+ \frac{f_1(z)}{r}+ \frac{f_2(z)}{r^2}+ \frac{f_3(z)}{r^3}+...,
\end{equation}
where each $f_i$ is itself a series expansion in positive powers of $z$. As in Ref.\ \cite{chrusciel1995gravitational, valiente1999logarithmic}, we restrict our attention to spacetimes where only finite powers of $z$ appear in the series, so that $f_i$ are polynomials in $z$.\footnote{Relaxing this condition would mean that the infinite series in $z$ that would appear in these calculations would in fact reduce to integer powers of $r$.  Therefore, our analysis, which treats the expansions in $1/r$ and $z$ independently, would no longer be valid.} Following Ref.\ \cite{valiente1999logarithmic}, we denote the degree of a polynomial $f$ as $\#f$.

Working with the Bondi definition of asymptotic flatness \cite{bondi1962gravitational,sachs1962gravitational},  we introduce Bondi coordinates $(u,r,x^{I})$ with $x^I=\{\theta,\phi\}$, such that the metric takes the form
\begin{equation} \label{metric}
ds^2=-Fe^{2\beta} du^2-2e^{2\beta}dudr+r^2h_{IJ} (dx^I-C^Idu)(dx^J-C^Jdu),
\end{equation}
where a residual gauge freedom in defining $r$ is fixed by imposing 
\begin{equation} \label{hdet}
 \det h = \det \omega
\end{equation}
with $\omega_{IJ}$ the standard metric on the round 2-sphere. This condition implies that $h_{IJ}$ has two degrees of freedom.

The Bondi definition of asymptotic flatness and the condition of polyhomogeneity requires that the metric parameters have the following large $r$ asymptotic form\footnote{Note the slight difference in notation in the expansion of $h_{IJ}$ compared to Refs.\ \cite{godazgar2019subleading, godazgar2019tower}, \textit{Cf.} equation (2.2) of Ref.\ \cite{godazgar2019subleading}.}
\begin{align}
\beta(u,r,x^I)  &=  \frac{\beta_0(z,u,x^I)}{r^2}+  \frac{\beta_1(z,u,x^I)}{r^3}+ \frac{\beta_2(z,u,x^I)}{r^4}  +o(r^{-4}), \notag \\[2mm]
F(u,r,x^I)  &=  1+\frac{F_0(z,u,x^I)}{r} +\frac{F_1(z,u,x^I)}{r^2} +\frac{F_2(z,u,x^I)}{r^3} +\frac{F_3(z,u,x^I)}{r^4}  +o(r^{-4}), \notag \\[2mm]
C^I(u,r,x^I)  &= \frac{C_0^I(z,u,x^I)}{r^2} +\frac{C_1^I(z,u,x^I)}{r^3}+\frac{C_2^I(z,u,x^I)}{r^4}+\frac{C_3^I(z,u,x^I)}{r^5}+o(r^{-5}), \label{metricexpansion} \\[2mm] \notag
h_{IJ}(u,r,x^I) &= \omega_{IJ}\hspace{-0.5mm} +\frac{B_{IJ}(z,u,x^I)}{r}+ \frac{\bar{C}_{IJ}(z,u,x^I)}{r^2} + \frac{\bar{D}_{IJ}(z,u,x^I)}{r^3} +\frac{\bar{E}_{IJ}(z,u,x^I)}{r^4}+o(r^{-4})
\end{align}
where
\begin{gather}
\bar{C}_{IJ}= C_{IJ}+\frac{1}{4}B^2\omega_{IJ}, \qquad
\bar{D}_{IJ}= D_{IJ}+\frac{1}{2}B_{KL}C^{KL}\omega_{IJ}, \notag \\[2mm]
\bar{E}_{IJ}= E_{IJ}+\bigg{(}\frac{1}{2}B_{KL}D^{KL}+\frac{1}{4}C^2-\frac{1}{32}(B^2)^2\bigg{)}\omega_{IJ} 
\end{gather}
with $B^2=B_{IJ}B^{IJ}$ and $C^2=C_{IJ}C^{IJ}$. This form of the $h_{IJ}$ expansion is chosen so that condition \eqref{hdet} simply translates to 
\begin{equation}
 \text{Tr} B= \text{Tr} C = \text{Tr} D= \text{Tr} E = 0
\end{equation}
at this order, where for some tensor $X_{IJ},$ $\Tr X \equiv \omega^{IJ} X_{IJ}$. Furthermore, the 2-sphere tensors $B,C,D$ and $E$ are independent and parameterise the two degrees of freedom of $h_{IJ}$ at each order.

\subsection{Asymptotic behaviour of Weyl scalar $\Psi_0$}
For spacetimes that are analytic in $1/r$, i.e.\ there exist no $\log$ terms, the Weyl tensor satisfies the so-called peeling property \cite{newman1962approach}, which can be simply stated as the fact that the Weyl tensor in the unphysical spacetime vanishes at null infinity.  In Newman-Penrose language \cite{newman1962approach}, this statement is equivalent to the fact that
\begin{equation}
 \Psi_i = O\left( \frac{1}{r^{5-i}}\right), \qquad i \in \{0,\ldots, 4\},
\end{equation}
where the $\Psi_i$ are the Newman-Penrose Weyl scalars defined with respect to a complex null frame $(\ell^a,n^a,m^a,\bar{m}^a)$,
\begin{gather}
 \Psi_0 = \ell^a m^b \ell^c m^d C_{abcd}, \quad \Psi_1 = \ell^a n^b \ell^c m^d C_{abcd}, \quad \Psi_2 = \ell^a m^b \bar{m}^c n^d C_{abcd}, \notag \\
 \Psi_3 = \ell^a n^b \bar{m}^c n^d C_{abcd}, \quad  \Psi_4 = n^a \bar{m}^b n^c \bar{m}^d C_{abcd}.
  \label{WeylScalars}
\end{gather}
As we shall explain below, the peeling property no longer holds in polyhomogeneous spacetimes \cite{chrusciel1995gravitational, Kroon:1998dv}.  Moreover, we shall find that the Weyl scalar $\Psi_0$ falls off too slowly.  This will lead us to make some further assumptions on the metric expansion \eqref{metricexpansion}.

We begin by choosing a complex null frame $e_\mu{}^a=(\ell^a,n^a,m^a,\bar{m}^a)$ with inverse $E^\mu{}_a$,
\begin{equation}
 g_{ab} = E^\mu{}_a E^\nu{}_b \ \eta_{\mu \nu}, \qquad \eta_{\mu \nu} = \text {{\footnotesize $ \begin{pmatrix}
                                                                       \begin{matrix} 0 & -1 \\ -1 & 0 \end{matrix} & \mathbf{0} \\
                                                                       \mathbf{0} & \begin{matrix} 0 & 1 \\ 1 & 0 \end{matrix}
                                                                    \end{pmatrix}$ }},
\end{equation}
where
\begin{align}
 \ell &= \frac{\partial}{\partial r}, \qquad n =  e^{- 2 \beta} \Bigg[ \frac{\partial}{\partial u} - \half F \frac{\partial}{\partial r} + C^I \frac{\partial}{\partial x^I} \Bigg], \qquad m = \frac{\hat{m}^I}{r}  \frac{\partial}{\partial x^I}, \notag\\
 \ell^\flat &= - e^{2\beta} du, \qquad n^{\flat} = - \Big( dr + \frac{1}{2} F du \Big), \qquad m^{\flat} = r\, \hat{m}_I\, (dx^I - C^I du),
 \label{AF:frame}
\end{align}
\begin{equation} \label{mmh}
 2 \hat{m}^{(I} \bar{\hat{m}}^{J)} = h^{IJ}
\end{equation}
with $h^{IJ}$ the matrix inverse of $h_{IJ}$.  The polyhomogeneous expansion \eqref{metricexpansion} implies that 
\begin{align}  \label{Psi0lowest}
\Psi_0 & \sim \frac{1}{r^3}\bigg{(}\mathcal{B}_{\theta\theta}-\frac{i}{\sin\theta} \mathcal{B}_{\theta\phi}\bigg{)}+\mathcal{O}(r^{-4}\log^{N_4}r),
\end{align}
where $\mathcal{B}_{IJ}=\partial_z B_{IJ}-\partial_z^2 B_{IJ}$.  Compare this with the fall-off of $\Psi_0$ in Refs. \cite{newman1968new,godazgar2019subleading, godazgar2019tower} 
\begin{equation}  
\Psi_0 = \frac{\Psi_0^5}{r^5}+\frac{\Psi_0^6}{r^6}+\frac{\Psi_0^7}{r^7}+o(r^{-7}).
\end{equation}

In this paper, in order to make progress, we will assume that $\Psi_0$ behaves asymptotically as $\mathcal{O}(r^{-4}\log^{N_4}r)$.  While, it is true that Newman-Penrose charges exist more generally for any polyhomogeneous spacetime defined by the fall-offs \eqref{metricexpansion} \cite{valiente1999logarithmic}, the analysis is much simpler if we assume that the leading order term in the shear of the null congruence defined by $\ell$ has no $\log$ terms \cite{kroon1998conserved}.  This is equivalent to the requirement that $\Psi_0 \sim \mathcal{O}(r^{-4}\log^{N_4}r)$, or that
\begin{equation}
 \mathcal{B}_{\theta\theta}-\frac{i}{\sin\theta} \mathcal{B}_{\theta\phi}=0,
\end{equation}
which is equivalent to the condition that $\mathcal{B}_{\theta\theta}=\mathcal{B}_{\theta\phi}=0$, given that $\mathcal{B}_{IJ}$ is real.  From the fact that $B_{IJ}$ is traceless and symmetric, we deduce that $\mathcal{B}_{IJ}$ is traceless and symmetric and hence the above condition is equivalent to
\begin{equation}
 \mathcal{B}_{IJ}=\partial_z B_{IJ}-\partial_z^2 B_{IJ}=0.
\end{equation}
 The fact that $B_{IJ}$ is a polynomial in $z$ of finite order, implies that 
 \begin{equation} \label{fshear}
  \partial_z B_{IJ}=0,
 \end{equation}
i.e.\ that $B_{IJ}$ is independent of $z$ and contains no $\log$ terms.  Henceforth, we shall assume that this condition always holds.  We shall find below that, together with this condition, the Einstein equations imply that all leading order terms in \eqref{metricexpansion} are independent of $z$.\footnote{At the next order, $\Psi_0\sim\frac{1}{r^4}\big{(}\mathcal{C}_{\theta\theta}+\frac{i}{\sin\theta} \mathcal{C}_{\theta\phi}\big{)}+\mathcal{O}(r^{-5}\log^{N_5}r)$ where $\mathcal{C}_{IJ}=-2C_{IJ}+3\partial_z C_{IJ}-\partial_z^2 C_{IJ}$.  Requiring that $\Psi_0=o(r^{-5})$ would imply that $C_{IJ}=0$, which recovers the fall-off conditions (2.2) in Ref.\ \cite{godazgar2019subleading}.}

\subsection{Notation}
For brevity, it will prove useful to use the following notation
\begin{equation}
 \int^{\lambda}X(z) \equiv e^{\lambda z} \int dz\, e^{-\lambda z} X(z),
\end{equation}
for $\lambda$ an integer, in order to reduce the size of some of the equations.  Furthermore, $\smallint^{\lambda}$ will be treated as an operator acting on the right, so we have for example
\begin{align}
\big{(}3\smallint^4-2\smallint^1+\smallint^0+6-\partial_z\big{)}X(z) = \hspace{1mm} 3e^{4z}\int  e^{-4z} X(z) dz & -2e^{z}\int  e^{-z} X(z) dz\notag \\
&+\int  X(z)dz +6X(z)-X^{\prime}(z).
\end{align}
For $\lambda\neq0$, $\int^{\lambda}$ does not change the order of the polynomial in $z$; see Appendix \ref{app:Polynomials}.  However, $\partial_z$ decreases the order by one and $\int^0$ increases it by one.

Moreover, angled brackets $\langle \ \rangle$ on pairs of indices will be used to denote the symmetric trace-free part; thus, for an arbitrary tensor $X_{IJ}$
\begin{equation}
 X_{\langle IJ \rangle} \equiv \frac{1}{2} \left( X_{IJ} + X_{JI} - \omega^{KL} X_{KL}\, \omega_{IJ} \right).
\end{equation}
For example, 
\begin{equation}
 B_{\langle I | K} {C^K}_{|J\rangle} = \frac{1}{2}B_{IK}{C^K}_J+\frac{1}{2}B_{JK}{C^K}_I-\frac{1}{2}B_{KL}C^{KL}\omega_{IJ}.
\end{equation}

\subsection{Einstein Equations}
We will assume that the energy-momentum tensor satisfies the fall-off conditions\footnote{Given some arbitrary vector $V_a$, we denote the components in the null basis as follows
\begin{equation*}
 \ell^a V_a \equiv V_0 = - V^1,\qquad n^a V_a \equiv V_1 = -V^0,\qquad m^a V_a \equiv V_m=V^{\bar{m}},
\end{equation*}
with the obvious generalisation to tensors.}
\begin{gather} 
T_{00}=o(r^{-4}), \quad T_{0m}=o(r^{-3}), \quad T_{01}=o(r^{-3}).
\end{gather}
The Einstein equation then yields
\begin{align} 
G_{00}=o(r^{-4}) \quad &\implies \quad \beta_0 = -\tfrac{1}{32} B^2, \\
G_{0m}=o(r^{-3}) \quad&\implies \quad C_0^I = -\tfrac{1}{2} D_J B^{IJ},\\
G_{01}=o(r^{-3}) \quad&\implies \quad  \partial_z F_0 =0,
\end{align}
where $D_I$ is the standard covariant derivative associated with the round-sphere metric $\omega_{IJ}$. Since $B_{IJ}$ is independent of $z$, we conclude the leading order terms in \eqref{metricexpansion} are all independent of $z$.

Assuming stronger fall-off conditions for the energy momentum tensor, the Einstein equations imply the following
\begin{align}
  G_{00} = o(r^{-5}) &\ \implies \ \beta_1 =  -\tfrac{1}{8} \big{(}\smallint^3+1\big{)} B_{IJ}C^{IJ},  \label{beta1}\\[1mm]
  G_{00} = o(r^{-6}) &\ \implies \ \beta_2 = -\tfrac{1}{8} \big{(}\smallint^4+1\big{)} B_{IJ}D^{IJ}-\tfrac{1}{8} \big{(}2\smallint^4+1\big{)}C^2+\tfrac{1}{16} \smallint^4\partial_z C_{IJ}\partial_z C^{IJ} \notag \\
&  \hspace{90mm}+\tfrac{1}{128} (B^2)^2,  \label{beta2} \\[1mm]
 G_{0m} = o(r^{-4}) &\ \implies \ C_1^I =  (C_1^0)^I-\tfrac{1}{3}\big{(}\smallint^3+2\smallint^0\big{)}D_J C^{IJ}, \label{C1}\\[1mm]
  G_{0m} = o(r^{-5}) &\ \implies \ \ C_2^I = -\tfrac{3}{4}B^{IJ}(C_1^0)_J -\tfrac{1}{3}\big{(}\smallint^4+2\smallint^1\big{)} D_J D^{IJ} -\tfrac{1}{2}\big{(}\smallint^4-\smallint^0\big{)}B^{IJ}D^K C_{JK} \notag\\
  &\hspace{20mm}+\tfrac{1}{4}\big{(}2\smallint^4-\smallint^3\big{)}B_{JK}D^I C^{JK}-\tfrac{1}{4}\smallint^3 C^{JK}D^I B_{JK} \notag\\
  &\hspace{20mm}  +\tfrac{1}{64}B^2D_J B^{IJ}-\tfrac{1}{16}B^{IJ}D_J B^2,
   \label{C2} \\[1mm]
  G_{0m} = o(r^{-6}) &\ \implies \ \ C_3^I =  \tfrac{9}{80}B^2(C_1^0)^I+3\smallint^5 (C_1^0)_J C^{IJ}-\tfrac{1}{3}\big{(}\smallint^5+2\smallint^2\big{)}D_J E^{IJ}\notag\\
 &\hspace{12mm} -\tfrac{1}{4}\smallint^4 D^{JK}D^I B_{JK}+\tfrac{1}{4}\big{(}2\smallint^5-\smallint^4\big{)}B_{JK}D^K D^{IJ}-\tfrac{1}{4}\big{(}\smallint^4-2\smallint^1\big{)}B^{IJ}D^K D_{JK} \notag\\
 &\hspace{20mm}+\tfrac{1}{2}\big{(}\smallint^5-\smallint^4\big{)}D^I C^2 -\tfrac{1}{24}\big{(}\smallint^5-3\smallint^4+2\smallint^2\big{)}D^I(\partial_z C^{JK} \partial_z C_{JK}) \notag \\
 &\hspace{20mm}-\tfrac{1}{6}\big{(}\smallint^5-\smallint^2\big{)}D^I C^{JK}\partial_z C_{JK}+\tfrac{2}{3}\big{(}\smallint^5-\smallint^2\big{)}\partial_z C^{IJ} D^K C_{JK} \notag \\
 & \hspace{20mm}  -\tfrac{2}{3}\big{(}\smallint^5-\smallint^2\big{)} C^{IJ} D^K C_{JK}-2\smallint^5\big{(} C^{IJ}\smallint dz D^K C_{JK} \big{)} \notag \\
& \hspace{20mm}+\tfrac{1}{8}\big{(}5\smallint^5+\smallint^3\big{)}B^{IJ}C^{KL}D_J B_{KL}-\tfrac{1}{8}\big{(}3\smallint^5-\smallint^3\big{)}B^{KL}C_{KL}D_J B^{IJ}\notag \\
& \hspace{20mm}+\tfrac{1}{8}\big{(}\smallint^5+\smallint^3\big{)}B^{IJ}B^{KL}D_J C_{KL}-\tfrac{1}{240}\big{(}7\smallint^5-40\smallint^2+18\smallint^0\big{)}B^2D_J C^{IJ}\notag\\
& \hspace{20mm}-\tfrac{1}{12}\big{(}5\smallint^5-2\smallint^2\big{)}C^{IJ}D_J B^2 +\tfrac{3}{160}B^2 D^I B^2, \label{C3} \\[1mm]
 G_{01} = o(r^{-4}) &\ \implies \ F_1 = -\tfrac{1}{2}D_I(C_1^0)^I-\tfrac{1}{3}\big{(}\smallint^3-\smallint^0\big{)}D_I D_J C^{IJ}\notag \\
&\hspace{20mm}+\tfrac{1}{2}D_IB^{IJ}D^K B_{JK}-\tfrac{1}{8}D_I B_{JK}D^I B^{JK}+\tfrac{3}{32}\big{(}\Box-2\big{)}B^2,  \label{F1} \\[1mm]
G_{01} = o(r^{-5}) &\ \implies \ F_2 = -\tfrac{3}{4}(C_1^0)^{I}D^JB_{IJ}-\tfrac{1}{3}\big{(}\smallint^4-\smallint^1\big{)}D_I D_J D^{IJ}\notag\\
&\hspace{20mm}+\tfrac{1}{4}\big{(}\smallint^4-\smallint^3\big{)}\Box\big{(}B_{IJ}C^{IJ}\big{)} -\tfrac{1}{4}\smallint^4C^{IJ}\Box B_{IJ}+\tfrac{1}{4}\big{(}2\smallint^4-\smallint^3\big{)}B_{IJ}C^{IJ} \notag\\
&\hspace{20mm} -\tfrac{1}{2}\big{(}\smallint^4-\smallint^0\big{)}D_IB^{IJ}D^K C_{JK}+\tfrac{1}{32}B^{IJ}B^{KL}D_KD_LB_{IJ} \notag\\
&\hspace{20mm} +\tfrac{1}{64}B^2D_ID_J B^{IJ}-\tfrac{1}{32}B^{IJ}D_IB_{KL}D_JB^{KL} \notag \\
&\hspace{20mm} +\tfrac{5}{64}D^I B^2 D^J B_{IJ},    \label{F2}
\end{align}
\begin{align}
   G_{01} = o(r^{-6}) &\ \implies \ F_3 = \tfrac{3}{4}(C_1^0)^I(C_1^0)_I +\tfrac{3}{160}D_I(B^2(C_1^0)^I)+\big{(}3\smallint^5-2\smallint^3-\smallint^0\big{)}(C_1^0)^ID^JC_{IJ}\notag \\
&\hspace{10mm}+\tfrac{3}{2}\big{(}2\smallint^5-\smallint^3\big{)} C_{IJ}D^J(C_1^0)^I -\tfrac{1}{3}\big{(}\smallint^5-\smallint^2\big{)}D_ID_JE^{IJ}+\tfrac{1}{4}\big{(}2\smallint^5-\smallint^4\big{)}B_{IJ}D^{IJ}\notag\\
&\hspace{10mm}+\tfrac{1}{4}\smallint^5B_{IJ}\Box D^{IJ}-\tfrac{1}{4}\smallint^4\Box(B_{IJ}D^{IJ})+\tfrac{1}{2}\smallint^1 D_IB_{JK}D^I D^{JK}\notag\\
&\hspace{10mm}+\tfrac{1}{2}\big{(}\smallint^5-\smallint^1\big{)}D_I B_{JK}D^J D^{IK}+\tfrac{1}{12}\big{(}4\smallint^5-6\smallint^4+3\smallint^3-\smallint^2\big{)}\Box C^2 \notag\\
&\hspace{10mm}-\tfrac{1}{24}\big{(}\smallint^5-3\smallint^4+3\smallint^3-\smallint^2\big{)}\Box\big{(}\partial_zC_{IJ}\partial_z C^{IJ}\big{)}+\tfrac{1}{12}\big{(}2\smallint^5-3\smallint^3+\smallint^2\big{)}\big{(}\partial_zC_{IJ}\Box C^{IJ}\big{)}\notag\\
&\hspace{10mm}-\tfrac{1}{2}\big{(}2\smallint^5-\smallint^3\big{)}\big{(}C_{IJ}\smallint dz\Box C^{IJ}\big{)}-\tfrac{1}{2}\big{(}2\smallint^5+\smallint^4-2\smallint^3\big{)}C^2\notag\\
&\hspace{10mm}+\tfrac{1}{8}\big{(}\smallint^4-\smallint^3\big{)}\big{(}\partial_zC_{IJ}\partial_z C^{IJ}\big{)} +\big{(}2\smallint^5-\smallint^3\big{)}\big{(}C_{IJ}\smallint dz C^{IJ}\big{)}\notag\\
&\hspace{10mm}-\tfrac{1}{12}\big{(}\smallint^5-\smallint^2\big{)}D_I C_{JK} D^I C^{JK}+\big{(}\smallint^5-\smallint^3\big{)}D^I C_{IJ} D_K C^{JK}\notag\\
&\hspace{10mm}-\tfrac{2}{3}\big{(}3\smallint^5-2\smallint^3\big{)}\big{(}D^I C_{IJ} \smallint dzD_K C^{JK}\big{)} +\tfrac{1}{3}\big{(}\smallint dz D^I C_{IJ} \big{)} \big{(}\smallint dz D_K C^{JK} \big{)}\notag\\
&\hspace{10mm}+\tfrac{1}{8}\big{(}5\smallint^5-3\smallint^3\big{)}B^{IJ}C^{KL}D_ID_J B_{KL}-\tfrac{1}{96}\big{(}40\smallint^5-33\smallint^3+8\smallint^2\big{)}C^{IJ}D_ID_JB^2\notag\\
&\hspace{10mm}-\tfrac{1}{8}\big{(}3\smallint^5-\smallint^3\big{)}B^{KL}C_{KL}D_I D_JB^{IJ}+\tfrac{1}{8}\big{(}\smallint^5-\smallint^3\big{)}B^{IJ}B^{KL}D_I D_J C_{KL}\notag\\
&\hspace{10mm}-\tfrac{1}{240}\big{(}7\smallint^5-30\smallint^3+20\smallint^2+3\smallint^0\big{)}B^2D_ID_JC^{IJ}\notag\\
&\hspace{10mm}-\tfrac{1}{240}\big{(}107\smallint^5-150\smallint^3+40\smallint^2+3\smallint^0\big{)}D_IC^{IJ}D_JB^2 \notag\\
&\hspace{10mm}+\tfrac{3}{4}\big{(}\smallint^5-\smallint^3\big{)}B^{IJ}D_IB_{KL}D_JC^{KL}-\tfrac{1}{4}\big{(}\smallint^5-\smallint^3\big{)}B^{KL}D_IB^{IJ}D_JC_{KL}\notag\\
&\hspace{10mm}-\tfrac{1}{8}\smallint^3C^{IJ}D_IB_{KL}D_J B^{KL}+\tfrac{1}{4}\big{(}\smallint^5+\smallint^3\big{)}C^{KL}D_IB_{KL}D_JB^{IJ}\notag\\
&\hspace{10mm}-\tfrac{1}{2}\smallint^3C_{KL}D_IB^{IK}D_{J}B^{JL}\notag\\
&\hspace{10mm}+\tfrac{5}{512}\big{(}B^2\big{)}^2-\tfrac{17}{5120}\Box\big{(}B^2\big{)}^2+\tfrac{13}{1024}D_IB^2 D^I B^2+\tfrac{3}{128}B^2 D_IB_{JK} D^{I}B^{JK}\notag\\
&\hspace{10mm}-\tfrac{1}{32}B^2 D_IB_{JK} D^J B^{IK},  \label{F3} \\[3mm]
  G_{mm} = o(r^{-3}) &\ \implies \ \partial_u C_{IJ} = 0,  \label{duCIJ}\\[3mm]
  G_{mm} = o(r^{-4}) &\ \implies \ \partial_u D_{IJ} = \tfrac{1}{8}B_{IJ}\partial_u B^2-\tfrac{1}{4}B_{IJ}F_0-\tfrac{1}{2}D_{\langle I}(C_1^0)_{J\rangle} \notag\\
 &\hspace{10mm} -\tfrac{1}{6}\big{(}4\smallint^3+2\smallint^0+3-3\partial_z\big{)}C_{IJ}+\tfrac{1}{6}\big{(}2\smallint^3+\smallint^0\big{)}\Box C_{IJ}-\tfrac{1}{8}B_{IJ}D_K D_LB^{KL} \notag\\
 &\hspace{10mm}+\tfrac{1}{32}D_{\langle I}D_{J\rangle}B^2+\tfrac{1}{2}D_{\langle I}\big{(}B_{J\rangle K}D_L B^{KL}\big{)}-\tfrac{1}{8}D_{\langle I|}B_{KL}D_{|J\rangle}B^{KL},   \label{duDIJ}
\end{align}
\begin{align}
  G_{mm} = o(r^{-5}) &\ \implies \ \partial_u E_{IJ} =\tfrac{1}{2}D^K\big{(}(C^0_1)_{\langle I}B_{J\rangle K}\big{)}+\tfrac{1}{2}\big{(}\smallint^3-1+\partial_z\big{)}C_{IJ}F_0 \notag\\
 &\hspace{10mm}+\tfrac{1}{2}B^{KL}C_{KL}\partial_u B_{IJ} + B_{K\langle I}C_{J\rangle L}\partial_u B^{KL} +\tfrac{1}{2}\smallint^3B_{IJ}C^{KL}\partial_u B_{KL} \notag\\
 &\hspace{10mm}-\tfrac{1}{4}\big{(}\smallint^3+1\big{)}C_{IJ}\partial_u B^2 +\tfrac{1}{3}\big{(}2\smallint^4+\smallint^1\big{)}D_{\langle I}D^K D_{J\rangle K} -\tfrac{1}{2}\big{(}2-\partial_z\big{)}D_{IJ} \notag\\
 &\hspace{10mm}-\tfrac{1}{2}\big{(}\smallint^3+1\big{)}C^{KL}D_KD_LB_{IJ}+\tfrac{1}{2}\big{(}\smallint^3+1\big{)}C^{KL}D_{\langle I}D_{J \rangle} B_{KL} \notag\\
 &\hspace{10mm}+\tfrac{1}{4}\big{(}\smallint^3+1\big{)}C_{IJ}D_K D_LB^{KL} -\tfrac{1}{6}\big{(}3\smallint^4-2\smallint^3-\smallint^0\big{)}B^{KL}D_{\langle I}D_{J\rangle}C_{KL} \notag\\
 &\hspace{10mm}+\tfrac{1}{6}\big{(}3\smallint^4-2\smallint^3-\smallint^0\big{)}B_{IJ}D_K D_LC^{KL}-\tfrac{1}{6}\big{(}3\smallint^4-\smallint^3+\smallint^0\big{)}B^{KL}D_K D_L C_{IJ} \notag\\
 &\hspace{10mm}-\smallint^4D_{\langle I}B^{KL}D_{J\rangle}C_{KL}+\tfrac{1}{2}\big{(}2\smallint^3+1\big{)}D_K B^{KL}D_LC_{IJ}+\smallint^4D_KC^{KL}D_L B_{IJ} \notag\\
 &\hspace{10mm}-\tfrac{1}{3}\big{(}3\smallint^4+2\smallint^3+\smallint^0\big{)}D^KC_{K\langle I}D^L B_{J\rangle L}+\tfrac{5}{32}D^K\big{(}B^2 D_{\langle I}B_{J\rangle K}\big{)} \notag\\
 &\hspace{10mm}-\tfrac{1}{8}D^K\big{(}B_{K\langle I}D_{J\rangle}B^2\big{)},     \label{duEIJ}
\end{align}
\begin{align}
  G_{11} = o(r^{-2}) &\ \implies \ \partial_u F_0 =  -\tfrac{1}{2}D_I D_J \partial_u B^{IJ}+\tfrac{1}{4}\partial_u B_{IJ} \partial_u B^{IJ}, \label{duF0} \\[2mm]
  G_{1m} = o(r^{-3}) &\ \implies \ \partial_u (C_1^0)^I =   \tfrac{1}{3}D^I F_0 +\tfrac{1}{6}\Box D_J B^{IJ}-\tfrac{1}{6}D^I D_J D_K B^{JK}+\tfrac{1}{8}B^{JK}\partial_u D^I B_{JK} \notag\\
 &\hspace{30mm}+\tfrac{5}{8}\partial_uB_{JK} D^IB^{JK}-\tfrac{2}{3}\partial_uB_{JK} D^K B^{IJ}-\tfrac{1}{6}D_J B^{IJ}, \label{duC10}
\end{align}
where $(C_1^0)^I$ has no $z$ dependence and $\Box=D^I D_I$ is the covariant Laplacian on the unit 2-sphere.

The above Einstein equations are the generalisations of the Einstein equations (2.15)--(2.26) of Ref.\ \cite{godazgar2019subleading}.  Setting $C_{IJ}=0$ and assuming that all tensors are $z$-independent, so that equation \eqref{app:cint} can be used, the above equations reduce to the respective equations in Ref.\ \cite{godazgar2019subleading} by taking $B_{IJ} \rightarrow C_{IJ}.$ 

Assuming the vacuum Einstein equations to the appropriate order, it is possible to deduce the $z$ order of each metric parameter in \eqref{metricexpansion}. In general,
\begin{gather}
 \#B_{IJ} = 0, \qquad \#C_{IJ} \equiv N_C\geq 0, \qquad  \#D_{IJ} \equiv N_D, \qquad \#E_{IJ} \equiv N_E,  \notag\\[2mm] 
 \#\beta_0 = 0, \qquad \#\beta_1 = N_C, \qquad \#\beta_2 = \max\{N_D,2N_C\},  \notag\\[2mm] 
 \#C_0^I = 0, \qquad \#C_1^I = N_C+1, \qquad \#C_2^I = N_D, \qquad \#C_3^I = \max\{N_E, 2N_C+1\},  \notag\\[2mm]
 \#F_0 = 0, \qquad \#F_1 = N_C+1, \qquad \#F_2 = N_D, \qquad \#F_3 = \max\{N_E, 2(N_C+1)\}.  
 \label{orders}
\end{gather} 
An important assumption that we shall rely upon in what follows is
\begin{equation}
N_E\geq N_D > N_C \geq 0.
\end{equation}
This is the case for generic initial data \cite{valiente1999logarithmic}. It is possible that in special cases, for example if $D_I C^{IJ}=0$, the above assumption does not hold. Nevertheless, all the charges obtained in this paper are still conserved in such cases.

\subsection{BMS group}
The asymptotic symmetry group of polyhomogeneous spacetimes is given by the BMS group, as with the smooth case \cite{chrusciel1995gravitational}.  This group is obtained by imposing that the variation of the metric under the generators of the asymptotic symmetry group respects the form of the metric and the gauge choices. These conditions imply a group of the form
\begin{equation*}
 \textup{BMS} = \textup{SL}(2,\mathbb{C}) \ltimes \textup{ST},
\end{equation*}
where $\textup{ST}$ represents the infinite affine group of supertranslations parameterised by a $u$ and $r$-independent function $s(x^I)$ and generated by diffeomorphisms of the form
\begin{equation} \label{BMSgen}
 \xi = s\, \partial_u +   \int dr \frac{e^{2\beta}}{r^2} h^{IJ} D_{J} s \  \partial_I - \frac{r}{2} \left( D_I \xi^I - C^I D
 _I s \right) \partial_r.
\end{equation}
As in Ref.\ \cite{godazgar2019subleading}, we shall concentrate on the supertranslation part of the BMS algebra.

We list below the variation of some of the metric components under supertranslations that will be useful later.
\begin{align}
\delta F_0 &=  s\partial_u F_0 -\tfrac{1}{2} \partial_u B^{IJ}D_{I}D_{J}s-D_J\partial_uB^{IJ}D_I s, \label{deltaF0} \\[2mm]
\delta (C_1^0)^I &= s\partial_u {(C_1^0)}^I+\tfrac{1}{16}\partial_uB^2D^Is +F_0D^Is-\tfrac{1}{4}B^{JK}D^I D_J D_Ks -\tfrac{1}{2}B^{IJ}D_J \Box s  \notag\\
 &\hspace{5mm}+\tfrac{1}{2}D^JB^{IK}D_J D_K s  -\tfrac{3}{4}D^IB^{JK}D_J D_Ks-\tfrac{1}{2} D_JB^{JK}D_K D^Is \notag\\
 &\hspace{5mm}-\tfrac{1}{2}D^I D^J B_{JK} D^Ks+\tfrac{1}{2}D^J D_KB^{IK} D_Js-B^{IJ} D_Js,  \label{deltaC10} \\[2mm]
\delta  B_{IJ} &= s\partial_uB_{IJ}-2 D_{\langle I} D_{J\rangle} s,  \label{deltaB} \\[2mm]
\delta  C_{IJ} &= s\partial_uC_{IJ},  \label{deltaC} 
\end{align}
\begin{align}
\delta  D_{IJ} &=s\partial_uD_{IJ}-2(C_1^0)_{\langle I}D_{J\rangle}s-\tfrac{1}{4}B_{IJ}B^{KL}D_K D_Ls-\tfrac{1}{8}B^2 D_{\langle I}D_{J\rangle}s+\tfrac{1}{8}D_{\langle I}B^2D_{J\rangle}s \notag\\
 &\hspace{5mm}+D_KB^{KL}B_{L\langle I}D_{J\rangle}s-\big{(}2\smallint^3+1\big{)}D^K C_{IJ}D_Ks+\tfrac{4}{3}\big{(}2\smallint^3+\smallint^0\big{)}D^KC_{K\langle I}D_{J\rangle}s \notag\\
 &\hspace{5mm}-\big{(}\smallint^3+1-\tfrac{1}{2}\partial_z\big{)}C_{IJ}\Box s,  \label{deltaD} \\[2mm]
\delta  E_{IJ} &= s\partial_uE_{IJ} +\tfrac{1}{2}(C_1^0)_{\langle I}B_{J\rangle K}D^K s -\tfrac{1}{2}\big{(}2\smallint^4+3-\partial_z\big{)}D_{IJ}\Box s\notag\\
 &\hspace{-2mm}+\tfrac{1}{3}\big{(}2\smallint^4+4\smallint^1-3\big{)}D^KD_{K\langle I}D_{J\rangle}s-\big{(}2\smallint^4+1\big{)}D_{\langle I} D_{J\rangle K}D^K s\notag\\
 &\hspace{-2mm}+\tfrac{1}{4}\big{(}4\smallint^3+2-\partial_z\big{)}B^{KL}C_{IJ}D_K D_L s-\tfrac{1}{2}\big{(}\smallint^3+1\big{)}\big{(}B_{IJ}C^{KL}D_K D_L s +B^{KL}C_{KL} D_{\langle I}D_{J\rangle} s\big{)} \notag\\
 &\hspace{-2mm}+\smallint^3\big{(}C_{KL}D_{\langle I}B^{KL}D_{J\rangle}s-C^{KL}D_K B_{IJ} D_L s\big{)}+C_{K \langle I} D_{J\rangle}B^{KL}D_L s +D_K B^{KL}C_{L\langle I}D_{J\rangle}s\notag\\
 &\hspace{-2mm}-\tfrac{1}{2}\partial_z\big{(}C_{IJ}D_K B^{KL} D_L s\big{)} -\big{(}\smallint^4-\smallint^3\big{)}B_{KL}D_{\langle I}C^{KL}D_{J\rangle}s-2\smallint^3B_{K\langle I}D_{J\rangle}C^{KL}D_{L}s\notag\\
 &\hspace{-2mm}+\tfrac{1}{2}B^{KL}D_K C_{IJ} D_L s+\tfrac{1}{3}\big{(}3\smallint^4+2\smallint^3+\smallint^0\big{)}D_{K}C^{KL}B_{L\langle I}D_{J\rangle}s +\tfrac{1}{3}\big{(}\smallint^3-\smallint^0\big{)}B_{IJ}D_KC^{KL}D_L s\notag\\
 &\hspace{-2mm} +\tfrac{5}{32}D^K\big{(}B^2B_{K\langle I}D_{J\rangle}s\big{)} +\tfrac{5}{32}B^2D^Ks D_{\langle I}B_{J\rangle K}-\tfrac{1}{8}B_{K\langle I}D_{J\rangle}B^2D^K s.  \label{deltaE}
\end{align}
These variations are guaranteed to preserve the form of the metric. However, we will impose further constraints on the metric via the Einstein equations by assuming particular fall-offs of the components of the energy-momentum tensor. If we impose a particular fall-off on one component, we may need to impose further conditions on other components so that the desired fall-off condition is preserved under the BMS action. The variation of a particular component (for fixed $\alpha, \beta \in \{0,1,m,\bar{m}\})$ is given by
\begin{equation} \label{Tfalloff}
\delta_{\xi} T_{\alpha \beta} = \xi^c \partial_c T_{\alpha\beta} +T_{c\beta}\partial_{\alpha }\xi^c +T_{c\alpha}\partial_{\beta }\xi^c.
\end{equation}
If we insist that $T_{\alpha \beta}=o(r^{-n})$, certain fall-off conditions must be obeyed by $T_{c\alpha}$ and $T_{c\beta}$. When assuming a particular fall-off condition, we will also assume that the relevant conditions are satisfied for the other components. This can always be done and presents no issues in our calculations.

\section{Standard BMS charges} \label{BMS}

The asymptotic charges associated with the asymptotic BMS symmetry group is given by the following expression \cite{barnich2002covariant} (see also Refs.\ \cite{iyer1994some, compere2018advanced})
\begin{equation} \label{BBexpression}
\slashed{\delta}\mathcal{Q}_\xi[ \delta g, g] = {\frac{1}{8\pi G}}  \int_S \star H[\xi,  g, \delta g] =  {\frac{1}{8\pi G}}  \int_S d\Omega \hspace{1mm} r^2 e^{2\beta} H^{ur} [\xi,  g, \delta g], 
\end{equation}
where we have used the form of the background metric of interest \eqref{metric} in the second equality. The 2-form $H$ is given by
\begin{align}
H= &\frac{1}{2} \Big\{ \xi_b g^{cd} \nabla_a \delta g_{cd} -\xi_b \nabla^c \delta g_{ac} +\xi^c  \nabla_b \delta g_{ca} \notag   \\[2mm]
&\hspace{20mm}  + \frac{1}{2} g^{cd} \delta g_{cd} \nabla_b \xi_a + \frac{1}{2} \delta g_{bc} (\nabla_a \xi^c - \nabla^c \xi_a) \Big\}dx^a \wedge dx^b.
\end{align}
The slash on the variational symbol $\delta$ in \eqref{BBexpression} signifies the fact that the variation is not, in general, integrable.

We have all the ingredients to compute charges, namely the background metric $g_{ab}$ given by \eqref{metric} and the symmetry generators given by \eqref{BMSgen}. Plugging the above into equation \eqref{BBexpression} leads to an expansion of the form \cite{godazgar2019subleading}
\begin{equation} \label{BMSexpansion}
 \ndelta \mathcal{Q}_\xi[\delta g, g]= \frac{1}{16 \pi G} \int_{S}\, d\Omega\ \Big\{ \ndelta \mathcal{I}_0 + \frac{\ndelta \mathcal{I}_1(z)}{r} + \frac{\ndelta \mathcal{I}_2(z)}{r^2} + \frac{\ndelta \mathcal{I}_3(z)}{r^3} + o(r^{-3}) \Big\},
\end{equation}
where each $\ndelta \mathcal{I}_i(z)$ is a polynomial of finite order in $z=\log r$. The first term $\ndelta \mathcal{I}_0$ in the expansion above has been derived previously for smooth asymptotically-flat spacetimes \cite{barnich2011bms}.  Below, we find that this result extends to polyhomogeneous spacetimes \cite{Chrusciel:1998he}.  Following Ref.\ \cite{godazgar2019subleading}, we extend the definition of BMS charges to subleading orders in a $1/r$-expansion.  Investigating these subleading BMS charges in the context of polyhomogeneous spacetimes is indeed the main aim of this paper.  We will find that the results in the polyhomogeneous case are analogous to those for smooth spacetimes, albeit, the expressions are rather more complicated.

\subsection{BMS charge at $\mathcal{O}(r^{0})$} \label{BMSr0}
At leading order, we find
\begin{equation} \label{BMSr0Charge}
\ndelta\mathcal{I}_0 = \delta(-2s F_0)+\frac{s}{2}\partial_uB_{IJ} \delta B^{IJ}.
\end{equation}
Observe that at this leading order in the variation of the BMS charges \eqref{BMSexpansion}, we do not encounter $\log r$ terms.  This is a direct consequence of the finite shear condition \eqref{fshear}, which implies that all leading order terms in the expansion \eqref{metricexpansion} are independent of $z$.

As in the smooth case \cite{barnich2011bms}, the non-integrability above is related to the existence of flux at infinity.  In particular, the charge is integrable if and only if $\partial_u B_{IJ}=0$, i.e.\ in the absence of Bondi news at null infinity \cite{wald2000general}.  The integrable part when integrated over the 2-sphere corresponds to leading-order BMS charges, which generalise the Bondi-Sachs 4-momentum corresponding to $s$  an $\ell=0$ or 1 spherical harmonic.

\subsection{BMS charge at $\mathcal{O}(r^{-1})$} \label{BMSr1}
At the next order, we obtain
\begin{align}
\slashed{\delta}\mathcal{I}_1 =&  s\delta\bigg{(} -2F_1-(1-\partial_z)D_I C_1^I+\tfrac{3}{16}(\Box-2)B^2+D_IB^{IJ}D^KB_{JK}-\tfrac{1}{4}D_I B_{JK} D^I B^{JK}\bigg{)}  \notag \\[2mm]
&\hspace{40mm}+\tfrac{1}{2}s\bigg{(}\partial_u B_{IJ}\delta C^{IJ} + \partial_uC_{IJ}\delta B^{IJ}\bigg{)}.
\end{align}
 If we assume the fall-off condition on the matter fields $T_{mm}=o(r^{-3})$, then from equations \eqref{duCIJ} and \eqref{deltaC}, $\delta C_{IJ} = s\partial_u C_{IJ} = 0$, so the non-integrable piece vanishes for all $s$. Assuming further that $T_{01}=o(r^{-4})$ and $T_{0m}=o(r^{-4})$, then equations \eqref{C1} and \eqref{F1} imply
\begin{equation}
\delta \mathcal{I}_1 = 0.
\end{equation}
Therefore, in this case $ \mathcal{I}_1 = 0 $ and there is no non-trivial charge. If, however, the fall-off of $T_{01}$ is weaker, we have non-vanishing charges given by the coefficients of the polynomial in $z$
\begin{equation} \label{I1Charges}
\mathcal{Q}_1(z) = \int_S d\Omega \bigg{(}-sT_{01}\bigg{|}_{r^{-4}} \bigg{)},
\end{equation}
provided that $T_{mm}=o(r^{-3})$ and $T_{0m}=o(r^{-4})$. It can be shown by considering \eqref{Tfalloff} that it is possible to have $T_{mm}=o(r^{-3})$ and $T_{0m}=o(r^{-4})$ with $T_{01}$ non-vanishing at this order. The higher order charges depend only on $C_{IJ}$. Since we have assumed $\partial_u C_{IJ}=0$, such terms are trivially conserved.  Therefore, the only interesting charge will be the one corresponding to the $z^0$ coefficient.

\subsection{BMS charge at $\mathcal{O}(r^{-2})$} \label{BMSr2}
Starting with weaker fall-off conditions $T_{00}=o(r^{-5})$, $T_{0m}=o(r^{-4})$ and $T_{mm}=o(r^{-3})$, which imply equations \eqref{beta1}, \eqref{C1} and \eqref{duCIJ}, the variation of the BMS charge at the next order is 
\begin{align}
\slashed{\delta}\mathcal{I}_2 &= s\delta \bigg{(}  -2F_2-\big{(}2-\partial_z\big{)}D_IC_2^I-3(C_1^0)^I D^J B_{IJ} -\tfrac{3}{2}D^J (C_1^0)^I B_{IJ} \notag \\[2mm]
&\hspace{10mm} -\tfrac{1}{4}\big{(}\smallint^3-2\smallint^0+1\big{)}B_{IJ}\Box C^{IJ}  -\tfrac{1}{4}\big{(}\smallint^3-1\big{)}C_{IJ}\Box B^{IJ} -\tfrac{1}{2}\big{(}\smallint^3+2\smallint^0\big{)}B_{IJ}C^{IJ} \notag \\[2mm]
&\hspace{10mm}-\tfrac{1}{2}\smallint^3 D_K B_{IJ}D^K C^{IJ} +2 \smallint^0 D_I B^{IJ} D^K C_{JK} +\tfrac{1}{8}B^2 D_I D_J B^{IJ}-\tfrac{1}{32}B^{IJ}D_I D_J B^2 \notag\\[2mm]
&\hspace{10mm}-\tfrac{1}{8}B^{IJ}D_I B_{KL} D_J B^{KL} +\tfrac{3}{16}D_I B^{IJ} D_J B^2 \bigg{)} \notag \\[2mm]
&\hspace{10mm}+s\bigg{(}\tfrac{1}{2}\delta B^{IJ}\partial_uD_{IJ} + \tfrac{1}{2}\partial_uB^{IJ} \delta D_{IJ}-\tfrac{1}{16}\partial_u B^2 \delta B^2 +\tfrac{1}{8}F_0 \delta B^2-\tfrac{1}{2}D^J (C_1^0)^I \delta B_{IJ} \notag \\[2mm]
&\hspace{10mm}-(C_1^0)^ID^J \delta B_{IJ}-\tfrac{1}{12}\big{(}2\smallint^3-2\smallint^0+3\partial_z\big{)}\delta B_{IJ}\Box C^{IJ} +\tfrac{1}{4}\big{(}2\smallint^3+2-\partial_z\big{)}C_{IJ}\Box \delta B^{IJ} \notag \\[2mm]
&\hspace{10mm}-\tfrac{1}{6}\big{(}4\smallint^3+2\smallint^0-3+3\partial_z\big{)}\delta B_{IJ}C^{IJ}+\tfrac{2}{3}\big{(}2\smallint^3+\smallint^0\big{)} D_K \delta B_{IJ}D^K C^{IJ} \notag \\[2mm]
&\hspace{10mm}+\tfrac{1}{2}\big{(}1-\partial_z\big{)}D_I \delta B^{IJ} D^K C_{JK} +\tfrac{1}{16}D_I D_J B^{IJ} \delta B^2 +\tfrac{1}{32}D_I D_JB^2 \delta B^{IJ} \notag \\[2mm]
&\hspace{10mm}+\tfrac{1}{16}D_I B^2 D_J \delta B^{IJ}+\tfrac{1}{2}B^{IJ}D^K B_{IK} D^L \delta B_{JL} +\tfrac{1}{8}\delta B^{IJ} D_I B_{KL}D_J  B^{KL} \bigg{)}.  
\end{align}
As ever, the above separation into the integrable and non-integrable parts is not unique. The choice above has been made in order to obtain the simplest expressions possible.  This will become most clear upon using further Einstein equations. If we further assume that $T_{0m}=o(r^{-5})$, $T_{01}=o(r^{-5})$ and $T_{mm}=o(r^{-4})$, which imply equations \eqref{C2}, \eqref{F2} and \eqref{duDIJ}, the above expression reduces to
\begin{align}
\slashed{\delta}\mathcal{I}_2 &= s D_I D_J \delta\bigg{(} -D^{IJ} +\tfrac{1}{16}B^2 B^{IJ}\bigg{)} \notag \\[2mm]
&\hspace{10mm}+s\bigg{(}\tfrac{1}{2}\delta B^{IJ}\partial_uD_{IJ} + \tfrac{1}{2}\partial_uB^{IJ} \delta D_{IJ}-\tfrac{1}{16}\partial_u B^2 \delta B^2 +\tfrac{1}{8}F_0 \delta B^2-\tfrac{1}{2}D^J (C_1^0)^I \delta B_{IJ} \notag \\[2mm]
&\hspace{10mm}-(C_1^0)^ID^J \delta B_{IJ}-\tfrac{1}{12}\big{(}2\smallint^3-2\smallint^0+3\partial_z\big{)}\delta B_{IJ}\Box C^{IJ} +\tfrac{1}{4}\big{(}2\smallint^3+2-\partial_z\big{)}C_{IJ}\Box \delta B^{IJ} \notag \\[2mm]
&\hspace{10mm} -\tfrac{1}{6}\big{(}4\smallint^3+2\smallint^0-3+3\partial_z\big{)}\delta B_{IJ}C^{IJ}+\tfrac{2}{3}\big{(}2\smallint^3+\smallint^0\big{)}D_K \delta B_{IJ}D^K C^{IJ} \notag \\[2mm]
&\hspace{10mm}+\tfrac{1}{2}\big{(}1-\partial_z\big{)}D_I \delta B^{IJ} D^K C_{JK} +\tfrac{1}{16}D_I D_J B^{IJ} \delta B^2 +\tfrac{1}{32} D_I D_JB^2 \delta B^{IJ} \notag \\[2mm]
&\hspace{10mm}+\tfrac{1}{16} D_I B^2 D_J \delta B^{IJ}+\tfrac{1}{2}B^{IJ}D^K B_{IK} D^L \delta B_{JL} +\tfrac{1}{8}\delta B^{IJ} D_I B_{KL} D_J  B^{KL} \bigg{)},  
\label{deltaI2Einstein}
\end{align}
where for brevity, we have not directly substituted the expression for $\partial_u D_{IJ}$. The integrable piece has $z$ degree $N_D>N_C\geq0$. A non-trivial charge could appear as a coefficient of each $z$ power in the integrable piece. We first consider the highest order---the coefficient of $z^{N_D}$. The non-integrable piece has maximum $z$ degree $N_C+1$ as can be seen from \eqref{orders}. If $N_D>N_C+1$ then each coefficient of $z^n$ for $n>N_C+1$ in the integrable piece gives a non-trivial charge. These are
\begin{equation} \label{I2Charges}
\mathcal{Q}_{2,n}^{\ell,m} =- {\frac{1}{16\pi G}} \int_S d\Omega \hspace{1mm} Y_{\ell m} \hspace{1mm}D_I D_J D^{IJ} \bigg{|}_{\mathcal{O}(z^n)} \hspace{5mm} \text{for} \hspace{5mm} n>N_C+1,
\end{equation}
where $Y_{\ell m}$ are spherical harmonics. However, inspecting equation \eqref{duDIJ}, the Einstein equation for $\partial_uD_{IJ}$, we notice that the right hand side has $z$ degree $N_C+1$; hence the higher order terms in $D_{IJ}$ do not evolve, i.e.\ they are constant in $u$. Therefore, the fact that the charges defined above are conserved is unsurprising.

The highest non-trivial order to consider is $\mathcal{O}(z^{N_C+1})$. We must calculate the coefficient of $z^{N_C+1}$ in the non-integrable piece and see what restrictions can be imposed in order to guarantee that this vanishes. The only terms in $\slashed{\delta}\mathcal{I}_2^{(non-int)}$ that have $z$ dependence are those containing $C_{IJ}$. Using equations \eqref{duDIJ}, \eqref{deltaB} and  \eqref{deltaD}, we obtain
\begin{align}
\slashed{\delta}\mathcal{I}_2^{(non-int)} \big{|}_{C_{IJ} \hspace{1mm}  \text{terms}} &= D^{\langle I}D^{J\rangle} s \bigg{(} -\tfrac{1}{6}\big{(}4\smallint^3+2\smallint^0+3-3\partial_z\big{)}sC_{IJ}+\tfrac{1}{6}\big{(}2\smallint^3+\smallint^0\big{)}s\Box C_{IJ}\notag \\[2mm]
&\hspace{10mm} -\big{(}2\smallint^3+1\big{)}D^K C_{IJ} D_K s +\tfrac{4}{3}\big{(}2\smallint^3+\smallint^0\big{)}D^K C_{IK}D_J s\notag \\[2mm]
&\hspace{10mm} -\tfrac{1}{2}\big{(}2\smallint^3+2-\partial_z\big{)} C_{IJ} \Box s \bigg{)}.
\label{I2Cterms}
\end{align}
Using the results of Appendix \ref{app:Polynomials}, we find that the terms of $z$ degree $N_C+1$ in the expression above are of the form
\begin{equation} \label{I2nonintNC1}
\slashed{\delta}\mathcal{I}_2^{(non-int)}  = \frac{1}{6}\int dz\hspace{1mm}   D^{\langle I}D^{J\rangle} s \bigg{(}  s(\Box-2) C_{IJ} +8D^K C_{K\langle I} D_{J\rangle} s \bigg{)} +  \mathcal{O}(z^{N_C}). 
\end{equation}
For any given $s(x)$, we can choose a $C_{IJ}(x)$ to make the expression in brackets an arbitrary symmetric traceless tensor. That is, for any traceless symmetric $X_{IJ}(x)$ and $s(x^I)$, we can find a  traceless symmetric solution $C_{IJ}(x)$ to the second order PDE
\begin{equation} \label{I2PDE}
s (\Box-2) C_{IJ} +8D^K C_{K\langle I}D_{J\rangle} s = X_{IJ}.
 \end{equation}
Since $C_{IJ}(z,u,x)$ can be freely specified on a Cauchy surface, the expression in \eqref{I2PDE} can be made arbitrary on the surface and so \eqref{I2nonintNC1} vanishes in general if and only if
 \begin{equation} \label{sl01}
 D_{\langle I}D_{J\rangle} s=0;
 \end{equation}
thus $s$ corresponds to an $\ell=0$ or 1 spherical harmonic (see Appendix \ref{app:sh}).

From \eqref{I2Cterms}, we observe that $\slashed{\delta}\mathcal{I}_2^{(non-int)} \big{|}_{C_{IJ} \hspace{1mm} \text{terms}} $ vanishes at all orders when $s$ obeys equation \eqref{sl01}. Moreover, from equation \eqref{HigherOrderMustDie} in Appendix \ref{app:Polynomials}, we have that for \eqref{I2Cterms} to vanish at a given order, it must vanish at all higher orders, in particular the highest order.  This means that $s$ must obey \eqref{sl01} for the coefficients of lower $z$ orders to be integrable. In conclusion, we deduce that at any order, \eqref{I2Cterms} vanishes if and only if $s$ is an $\ell=0$ or 1 spherical harmonic.

Assuming equation \eqref{sl01}, the non-integrable part of equation \eqref{deltaI2Einstein} reduces to
\begin{equation}
\slashed{\delta}\mathcal{I}_2^{(non-int)} = D^I\big{(} -s^2(C_1^0)^J \partial_u B_{IJ} +\tfrac{1}{16}s^2D^JB^2 \partial_u B_{IJ} +\tfrac{1}{2} s^2 B^{JK}D^L B_{KL} \partial_u B_{IJ}\big{)}, 
\end{equation}
 which is a total derivative and can, therefore, be ignored. Hence, at all orders in $z$ we obtain the (unintegrated) charges
 \begin{equation}
\mathcal{I}_2 = s D_I D_J \big{(} -D^{IJ} +\tfrac{1}{16}B^2 B^{IJ}\big{)}.
\end{equation}
However, up to total derivatives, this is equal to
  \begin{align}
\mathcal{I}_2 &=  D_I D_Js \big{(} -D^{IJ} +\tfrac{1}{16}B^2 B^{IJ}\big{)} \notag \\[2mm]
&=  D_{\langle I} D_{J\rangle} s \big{(} -D^{IJ} +\tfrac{1}{16} B^2 B^{IJ}\big{)}\\[2mm]
&=  0,
\end{align}
 where in the second line we have used the fact that $B_{IJ}$ and $D_{IJ}$ are trace-free and in the third line we have used equation \eqref{sl01}. Therefore, the only non-trivial charges obtained at $\mathcal{O}(r^{-2})$ are those given in equation \eqref{I2Charges}.

\subsection{BMS charge at $\mathcal{O}(r^{-3})$} \label{BMSr3}
Starting with the previous fall-off conditions $T_{00}=o(r^{-5})$, $T_{0m}=o(r^{-5})$, $T_{mm}=o(r^{-4})$ and $T_{01}=o(r^{-5})$, which imply equations \eqref{beta1}, \eqref{C1}, \eqref{C2}, \eqref{duCIJ} and \eqref{duDIJ}, we find that, to leading orders in $z$,
\begin{align}
\slashed{\delta}\mathcal{I}_3 &= s\delta \bigg{(}  -2F_3+4\beta_2+2\Box \beta_2-\big{(}3-\partial_z\big{)}D_IC_3^I-2\smallint^0 (C_1^0)^ID^J C_{IJ}   \notag\\[2mm]
&\hspace{10mm} +\tfrac{1}{2}\smallint^1 B^{IJ}\Box D_{IJ}+ \tfrac{1}{2}D_{IJ}\Box B^{IJ}+\tfrac{1}{2}D_KD_{IJ}D^K B^{IJ}+2\smallint^1 D_I B^{IJ}D^K D_{JK} \notag\\[2mm]
&\hspace{10mm} -\tfrac{1}{2}\big{(}2\smallint^1-1\big{)}B^{IJ}D_{IJ}-\tfrac{1}{4}\smallint^0 D^I(B^2D^J C_{IJ})\bigg{)} \notag\\[2mm]
&+s\bigg{(}\tfrac{1}{2}\delta B^{IJ}\partial_uE_{IJ} + \tfrac{1}{2}\partial_uB^{IJ} \delta E_{IJ} -\tfrac{1}{12}\big{(}2\smallint^4-2\smallint^1-3+3\partial_z\big{)}\delta B^{IJ}\Box D_{IJ}\notag\\[2mm]
&+\tfrac{1}{4}\big{(}2\smallint^4+3-\partial_z\big{)}D_{IJ}\Box \delta B^{IJ}+\tfrac{1}{2}\big{(}2-\partial_z\big{)}D_K D_{IJ} D^K \delta B^{IJ} +\tfrac{2}{3}\big{(}2\smallint^4+\smallint^1\big{)}D_I\delta B^{IJ} D^K D_{JK}\notag\\[2mm]
& -\tfrac{1}{6}\big{(}4\smallint^4+2\smallint^1-6+3\partial_z\big{)}D_{IJ}\delta B^{IJ}+\smallint^0\big{[}\tfrac{5}{6}D^I\delta B^2 D^J C_{IJ} +\tfrac{3}{4}\delta B^2 D^I D^J C_{IJ}\notag\\[2mm]
&-\tfrac{4}{3}B^{IJ}\delta {B_{I}}^K D_J D_L {C_K}^L -\tfrac{3}{2} {B_{I}}^K \delta B^{IJ} D_J D_L {C_K}^L-2B^{IJ}D_I\delta B_{JK}D_L C^{KL}\notag\\[2mm]
&-\tfrac{3}{2}\delta B^{IJ}D_I B_{JK}D_L C^{KL}    -\tfrac{4}{3}\delta B_{IJ}D_K B^{IK}D_L C^{JL} -\tfrac{11}{6}B_{IJ}D_K\delta B^{IK}D_L C^{JL}\notag\\[2mm]
&-\tfrac{1}{6}\delta B^{IJ} D^K B_{IJ} D^L C_{KL}+\tfrac{1}{6} B^{IJ} D^K \delta B_{IJ} D^L C_{KL} \big{]}\bigg{)}   +\mathcal{O}\big{(}z^{N_C}\big{)}.
\end{align}
As before, it should be emphasised that the separation into integrable and non-integrable pieces is not unique and the form above has been chosen to make the following expressions simpler.

Imposing a stronger fall-off of the energy-momentum tensor, $T_{00}=o(r^{-6})$, $T_{01}=o(r^{-6})$, $T_{0m}=o(r^{-6})$ and $T_{mm}=o(r^{-5})$ implies equations \eqref{beta2}, \eqref{F3}, \eqref{C3} and \eqref{duEIJ}. Substituting these equations into the expressions above, in the integrable piece, terms will appear that depend only on $C_{IJ}$.  We drop such terms, since $\delta C_{IJ}=0$ and one can add any arbitrary term depending only on $C_{IJ}$ to the conserved charges. After applying some Schouten identities (See Appendix \ref{app:Schouten}), up to total derivatives
\begin{align}
\slashed{\delta}\mathcal{I}_3 &= s\delta \bigg{(}  -D_I D_J E^{IJ} +\mathcal{O}(z^{N_C})\bigg{)}\notag\\[2mm]
&+s\bigg{(}\tfrac{1}{2}\delta B^{IJ}\partial_uE_{IJ} +\tfrac{1}{2}\partial_uB^{IJ} \delta E_{IJ} -\tfrac{1}{12}\big{(}2\smallint^4-2\smallint^1-3+3\partial_z\big{)}\delta B^{IJ}\Box D_{IJ}\notag\\[2mm]
&+\tfrac{1}{4}\big{(}2\smallint^4+3-\partial_z\big{)}D_{IJ}\Box \delta B^{IJ}+\tfrac{1}{2}\big{(}2-\partial_z\big{)}D_K D_{IJ} D^K \delta B^{IJ} +\tfrac{2}{3}\big{(}2\smallint^4+\smallint^1\big{)}D_I\delta B^{IJ} D^K D_{JK}\notag\\[2mm]
& -\tfrac{1}{6}\big{(}4\smallint^4+2\smallint^1-6+3\partial_z\big{)}D_{IJ}\delta B^{IJ}+\smallint^0\big{[}\tfrac{5}{6}D^I\delta B^2 D^J C_{IJ} +\tfrac{3}{4}\delta B^2 D^I D^J C_{IJ}\notag\\[2mm]
& -\tfrac{4}{3}B^{IJ}\delta {B_{I}}^K D_J D_L {C_K}^L -\tfrac{3}{2} {B_{I}}^K \delta B^{IJ} D_J D_L {C_K}^L-2B^{IJ} D_I\delta B_{JK} D_L C^{KL}\notag\\[2mm]
&-\tfrac{3}{2}\delta B^{IJ}D_I B_{JK} D_L C^{KL}  -\tfrac{4}{3}\delta B_{IJ}D_K B^{IK}D_L C^{JL} -\tfrac{11}{6}B_{IJ}D_K\delta B^{IK}D_L C^{JL}\notag\\[2mm]
&-\tfrac{1}{6}\delta B^{IJ} D^K B_{IJ} D^L C_{KL} +\tfrac{1}{6} B^{IJ} D^K \delta B_{IJ} D^L C_{KL} \big{]} +\mathcal{O}\big{(}z^{N_C}\big{)}\bigg{)},  
\label{I3Einstein}
\end{align}
where for brevity, we have not yet substituted in the expression for $\partial_u E_{IJ}$. The integrable piece has $z$ degree $N_E$ where $N_E\geq N_D > N_C$. Using the appropriate Einstein equations and the metric variations shows that the non-integrable piece has $z$ degree $N_D$. If $N_E>N_D$, terms $\mathcal{O}(z^{N_D+1})$ or higher are integrable and we have charges
\begin{equation} \label{I3BoringCharge}
\mathcal{Q}_{3,n}^{\ell, m} =- {\frac{1}{16\pi G}} \int_S d\Omega \hspace{1mm} Y_{\ell m} \hspace{1mm}D_I D_J E^{IJ} \bigg{|}_{\mathcal{O}(z^n)} \hspace{5mm} \text{for} \hspace{5mm} n>N_D.
\end{equation}
As with charges \eqref{I2Charges} derived in Section \ref{BMSr2}, the existence of such conserved charges is unsurprising, since we observe from Einstein equation \eqref{duEIJ} that terms of this degree in $E_{IJ}$ are constant in $u$. The highest order at which the existence of a charge is not immediately obvious is at $\mathcal{O}(z^{N_D})$. After applying the metric variations to \eqref{I3Einstein} and tidying up the resulting expression using Schouten identities (see Appendix \ref{app:Schouten}), we find all $\partial_u$ terms at this order result in a total derivative and can hence be ignored.  Recalling that $N_C<N_D$, it is possible that $N_C+1=N_D$ so the remaining terms in $\slashed{\delta}\mathcal{I}_3^{(non-int)} $ that can contribute to this order arise from $C_{IJ}$ and $D_{IJ}$ terms. We have
\begin{equation}
\slashed{\delta}\mathcal{I}_3^{(non-int)} = \hspace{1mm}\bigg{(}\slashed{\delta}\mathcal{I}_3^{(non-int)}\big{|}_{D \hspace{1mm} \text{terms}}+\slashed{\delta}\mathcal{I}_3^{(non-int)}\big{|}_{BC \hspace{1mm} \text{terms}}\bigg{)}+ \mathcal{O}(z^{N_C}), 
\end{equation}
where 
\begin{align}
\slashed{\delta}\mathcal{I}_3^{(non-int)}\big{|}_{D \hspace{1mm} \text{terms}} &= s\bigg{(} -\tfrac{1}{2}\big{(}\smallint^1+1-\partial_z\big{)}\Box D^{IJ}D_I D_J s   +\tfrac{1}{2}\big{(}4\smallint^4+2\smallint^1-2+\partial_z\big{)}D^{IJ} D_I D_J s \notag\\[2mm]
 &\hspace{10mm}-\tfrac{1}{2}\big{(}2\smallint^4+3-\partial_z\big{)}D^{IJ}\Box D_I D_J s-\big{(}2-\partial_z\big{)}D^K D^{IJ} D_K D_I D_J s\notag\\[2mm]
 &\hspace{10mm}-\tfrac{2}{3}\big{(}2\smallint^4+\smallint^1\big{)}D_I D^{IJ} D_J \Box s -\tfrac{4}{3}\big{(}2\smallint^4+\smallint^1\big{)}D_I D^{IJ} D_J s \bigg{)}
 \label{I3D}
\end{align}
and
\begin{align} 
\slashed{\delta}\mathcal{I}_3^{(non-int)}\big{|}_{BC \hspace{1mm} \text{terms}}&=\frac{1}{3}\int dz \bigg{(} B^{IJ}D_L C^{KL} D_K D_I D_J s -\tfrac{1}{2} B^{IJ}D_K {C_I}^K D_J \Box s \notag\\[2mm]
&\hspace{20mm} -B^{IJ}D_K {C_I}^K D_J  s \bigg{)}+ \mathcal{O}(z^{N_C}). 
\label{I3BC}
\end{align}
There is no Einstein equation relating $D_{IJ}$ and $C_{IJ}$, so the contributions from the two terms above, namely $\slashed{\delta}\mathcal{I}_3^{(non-int)}\big{|}_{D \hspace{1mm} \text{terms}}$ and $\slashed{\delta}\mathcal{I}_3^{(non-int)}\big{|}_{BC \hspace{1mm} \text{terms}}$ need to vanish independently in \eqref{I3Einstein} in order for the charge to be integrable in general. We focus on $\slashed{\delta}\mathcal{I}_3^{(non-int)}\big{|}_{D \hspace{1mm} \text{terms}}$ to begin with. The $\mathcal{O}(z^{N_D})$ coefficients can be calculated using \eqref{gettingcoefficients} in Appendix \ref{app:Polynomials}. We find that
\begin{align}
\slashed{\delta}\mathcal{I}_3^{(non-int)}\big{|}_{D \hspace{1mm} \text{terms}} &= s\bigg{(} -\tfrac{5}{2} D^{IJ}D_I D_J s -\tfrac{5}{4} D^{IJ} \Box D_I D_J s - 2 D^K D^{IJ} D_K D_I D_J s\notag\\[2mm]
& +D_I D^{IJ} D_J \Box s +2 D_I D^{IJ} D_J s\bigg{)}\bigg{|}_{\mathcal{O}(z^{N_D})} + \mathcal{O}(z^{N_D-1}).
\label{I3DHighest}
\end{align} 
Now, in order to simplify this expression, we add to it an additional term
\begin{equation} \label{iszero}
5s(D^{IJ}D_I D_J s +\tfrac{1}{4} D^{IJ}D_ID_J \Box s -\tfrac{1}{4}D^{IJ} \Box D_I D_J s),
\end{equation}
which vanishes upon use of the the Ricci identity and the fact $D_{IJ}$ is traceless; thus we have not changed the non-integrable piece \eqref{I3DHighest}, which becomes
\begin{align}
&\slashed{\delta}\mathcal{I}_3^{(non-int)}\big{|}_{D \hspace{1mm} \text{terms}} = s\bigg{(} -\tfrac{5}{2}D^{IJ} \Box D_I D_J s+\tfrac{5}{4} D^{IJ} D_I D_J \Box s+\tfrac{5}{2} D^{IJ}D_I D_Js \notag\\[2mm]
&\hspace{10mm}-2D^K D^{IJ} D_K D_I D_J s +D_I D^{IJ} D_J \Box s +2D_I D^{IJ} D_J s\bigg{)} \bigg{|}_{\mathcal{O}(z^{N_D})} + \mathcal{O}(z^{N_D-1}).
\label{afteridentity}
\end{align}
Up to total derivatives, the first line in \eqref{afteridentity} can be written as
\begin{align}
&\tfrac{5}{2}sD^K D^{IJ} D_K D_I D_J s +\tfrac{5}{2} D^{IJ}D^K s  D_K D_I D_J s  -\tfrac{5}{4} s D_I D^{IJ} D_J \Box s -\tfrac{5}{4}  D^{IJ} D_I s D_J \Box s \notag\\[2mm]
 &\hspace{20mm}-\tfrac{5}{2} s D_I D^{IJ} D_J s-\tfrac{5}{2} D^{IJ}D_I s D_J s,
\end{align}
so that
\begin{align}
&\slashed{\delta}\mathcal{I}_3^{(non-int)}\big{|}_{D \hspace{1mm} \text{terms}} = \bigg{(} \tfrac{1}{2} sD^K D^{IJ} D_K D_I D_J s + \tfrac{5}{2}D^{IJ}D^K s  D_K D_I D_J s  \notag\\[2mm]
&\hspace{30mm}   - \tfrac{1}{4} s D_I D^{IJ} D_J \Box s - \tfrac{5}{4} D^{IJ} D_I s D_J \Box s  \notag\\[2mm]
&\hspace{30mm}  - \tfrac{1}{2} s D_I D^{IJ} D_J s- \tfrac{5}{2}D^{IJ} D_I s D_J s \bigg{)} \bigg{|}_{\mathcal{O}(z^{N_D})} + \mathcal{O}(z^{N_D-1})  \notag\\[2mm]
&=  \tfrac{1}{4}\big{ (}sD^K D^{IJ} + 5D^{IJ} D^K s\big{)} \big{(}2 D_K D_I D_J s -\omega_{KI}D_J \Box s -2\omega_{IK}D_J s\big{)} \bigg{|}_{\mathcal{O}(z^{N_D})} + \mathcal{O}(z^{N_D-1}). 
\end{align}
The factor $\big{ (}sD_K D_{IJ} + 5D_{IJ} D_K s\big{)}$ is an arbitrary tensor that is symmetric and trace-free on its $I, J$ indices. Thus, for the $\mathcal{O}(z^{N_D})$ term to vanish for general $D_{IJ}$, it is necessary that the projection onto the traceless, symmetric part of the other factor vanishes, i.e.
\begin{equation} \label{I3sequation}
2D_K D_{\langle I}D_{J \rangle} s - \omega_{K\langle I} D_{J \rangle} \Box s -2\omega_{K\langle I} D_{J \rangle}  s =0.
\end{equation}
In Appendix \ref{app:sh}, we show that the above equation \eqref{I3sequation} is satisfied if and only if $s$ is a superposition of $\ell=0,1$ and $2$ spherical harmonics. As with $\slashed{\delta}\mathcal{I}_2$ , there is little point considering $\ell=0$ and $1$ modes since their contribution to the integrable piece is trivially zero; hence we now consider exclusively the case in which $s$ is an $\ell=2$ spherical harmonic, which means in particular that
\begin{equation} \label{lis2}
(\Box+6)s=0.
\end{equation}
Combining \eqref{lis2} with \eqref{I3sequation}, we have
\begin{align} \label{lis2other}
D_K D_{\langle I} D_{J\rangle} s -\tfrac{1}{3}\omega_{K\langle I}D_{J\rangle} \Box s=0 \hspace{5mm} \text{and} \hspace{5mm} D_K D_{\langle I} D_{J\rangle} s +2 \omega_{K\langle I}D_{J\rangle}  s=0.
\end{align}

Next, we consider the terms in $\slashed{\delta}\mathcal{I}_3^{(non-int)}\big{|}_{BC \hspace{1mm} \text{terms}}$. Reorganising these terms, equation \eqref{I3BC} becomes
\begin{align}
\slashed{\delta}\mathcal{I}_3^{(non-int)}\big{|}_{BC \hspace{1mm} \text{terms}}   &= \frac{1}{3} \int dz \bigg{\{} B^{IJ}D_L C^{KL} \big{(}D_K D_I D_J -\tfrac{1}{3} \omega_{KI}D_J \Box s\big{)} \notag\\[2mm]
&\hspace{20mm} -\tfrac{1}{6} B^{IJ}D_K {C_I}^K D_J (\Box+6)s\bigg{\}} + \mathcal{O}(z^{N_C}).
\end{align}
Given that $B_{IJ}$ is trace-free and symmetric, we can use equations \eqref{lis2other} and \eqref{lis2} to conclude that
\begin{align}
\slashed{\delta}\mathcal{I}_3^{(non-int)}\big{|}_{BC \hspace{1mm} \text{terms}}   &= 0.
\end{align}
Therefore, even in the extreme case $N_D = N_C+1$, the $\mathcal{O}(z^{N_D})$ term still vanishes for an $\ell=2$ spherical harmonic. The fact that the non-integrable charges vanish at $z$ degree $N_C+1$ means that we have 5 conserved charges
\begin{equation} \label{I3Charge}
\mathcal{Q}_m =- {\frac{1}{16\pi G}} \int_S d\Omega \hspace{1mm} Y_{2m} \hspace{1mm} D_I D_J E^{IJ} \bigg{|}_{\mathcal{O}(z^{N_D})} \hspace{5mm} \text{for} \hspace{2mm} m=0, \pm 1, \pm 2.
\end{equation}
We shall show below in Section \ref{NP} that these charges correspond to half of the set of Newman-Penrose charges that exist in such polyhomogeneous spacetimes \cite{kroon1998conserved}.

We now consider the non-integrable piece at lower orders. Using the result in \eqref{HigherOrderMustDie}, for the non-integrable piece to vanish at lower orders for general $D_{IJ}$, it is necessary that it vanishes at the highest order for general $D_{IJ}$ and so $s$ must be an $\ell=2$ spherical harmonic. In this case, applying equation \eqref{PXz} to the expression in \eqref{I3D}, the contribution from $D_{IJ}$ terms at $\mathcal{O}(z^{N_D-1})$ is
\begin{align}
\slashed{\delta}\mathcal{I}_3^{(non-int)}\big{|}_{D \hspace{1mm} \text{terms}} &= N_D\, s\big{(} \Box D^{IJ}D_I D_J s - \tfrac{5}{8} D^{IJ}D_I D_J s+\tfrac{9}{16} D^{IJ}\Box D_I D_J s \notag\\[2mm]
&+D^K D^{IJ} D_K D_I D_J s +\tfrac{3}{4} D_I D^{IJ} D_J \Box s +\tfrac{3}{2} D_I D^{IJ} D_J s\big{)} \bigg{|}_{\mathcal{O}(z^{N_D-1})} \notag\\[2mm]
&\hspace{80mm} + \mathcal{O}(z^{N_D-2}).
\end{align}
Up to total derivatives, this becomes
\begin{align}
\slashed{\delta}\mathcal{I}_3^{(non-int)}\big{|}_{D \hspace{1mm} \text{terms}} &= N_D\, D^{IJ}\big{(} \tfrac{9}{16}s\Box D_I D_J s -\tfrac{3}{4}s D_I D_J \Box s +D^K s D_K D_I D_J s -\tfrac{3}{4} D_I\Box s D_Js \notag\\[2mm]
&+\Box s D_I D_J s -\tfrac{17}{8}s D_I D_J s -\tfrac{3}{2} D_I s D_J s \big{)} \bigg{|}_{\mathcal{O}(z^{N_D-1})} + \mathcal{O}(z^{N_D-2}). 
\end{align}
Then using equations \eqref{lis2} and \eqref{lis2other}, this reduces to
\begin{equation} \label{I3obstruction}
\slashed{\delta}\mathcal{I}_3^{(non-int)}\big{|}_{D \hspace{1mm} \text{terms}} = N_D\, D^{IJ}\big{(} -\tfrac{19}{4}sD_I D_J s +D_I s D_J s \big{)} \bigg{|}_{\mathcal{O}(z^{N_D-1})} + \mathcal{O}(z^{N_D-2}),
\end{equation}
which for general $D_{IJ}$ and $s$ an $\ell=2$ spherical harmonic is not zero. Any further restriction on $s$ will make the integrable piece vanish. There is no need to check $\slashed{\delta}\mathcal{I}_3^{(non-int)}\big{|}_{BC \hspace{1mm} \text{terms}}$ at this order since there is no equation linking $D_{IJ}$ to $C_{IJ}$ and $B_{IJ}$ that could result in a cancellation in the non-integrable piece. We deduce that $\slashed{\delta}\mathcal{I}_3^{(non-int)}\big{|}_{D \hspace{1mm} \text{terms}}$ is non-vanishing at this order and hence there are no charges at this order, nor subsequent orders as implied by \eqref{HigherOrderMustDie}. 

In summary, the complete set of conserved charges obtained at $\mathcal{O}(r^{-3})$ are given by \eqref{I3BoringCharge} and \eqref{I3Charge}.

\section{Dual BMS charges} \label{Dual}
We now turn to the tower of dual charges defined in Ref.\ \cite{godazgar2019tower}, given by the expression
\begin{equation} \label{DualCharge}
\slashed{\delta}\mathcal{\widetilde{Q}}_\xi[ \delta g, g] = {\frac{1}{8\pi G}}  \int_S  \widetilde{H}[\xi,  g, \delta g] =  {\frac{1}{8\pi G}}  \int_S d\Omega \hspace{1mm} \frac{\widetilde{H}_{\theta \phi} [\xi,  g, \delta g] }{\sin \theta},
\end{equation}
where we have used the form of the background metric of interest \eqref{metric} in the second equality with the 2-form $\widetilde{H}$ given by
\begin{align} \label{tildeH}
\widetilde{H}= &\frac{1}{4} \, \delta g_{bc} (\nabla_a \xi^c+\nabla^c \xi_a)  \, dx^a \wedge dx^b.
\end{align}
The dual BMS charges can be derived from first principles from the Palatini-Holst action \cite{Godazgar:2020gqd, Godazgar:2020kqd}.  We will consider a $1/r$-expansion of the variation of the dual BMS charge
\begin{equation} \label{Dualexpansion}
 \ndelta \mathcal{\widetilde{Q}}_\xi[\delta g, g]= \frac{1}{16 \pi G} \int_{S}\, d\Omega\ \Big\{ \ndelta \mathcal{\widetilde{I}}_0 + \frac{\ndelta \mathcal{\widetilde{I}}_1(z)}{r} + \frac{\ndelta \mathcal{\widetilde{I}}_2(z)}{r^2} + \frac{\ndelta \mathcal{\widetilde{I}}_3(z)}{r^3} + o(r^{-3}) \Big\}.
\end{equation}
The calculations will be analogous to those in Section \ref{BMS} with similar results being obtained, as with the smooth case \cite{godazgar2019tower}.

Following Ref.\ \cite{godazgar2019new}, it will be useful to define the twist of a symmetric tensor $X_{IJ}$
\begin{equation} \label{Xtwist}
\widetilde{X}^{IJ} = X_K{}^{(I} \epsilon^{J)K}, \qquad \epsilon_{IJ} =  
\begin{pmatrix}                                                                                0 & 1 \\ -1 & 0                                                              \end{pmatrix} \sin \theta.
\end{equation}
Note, if $X^{IJ}$ is trace-free, we can drop the symmetrisation in the definition \eqref{Xtwist}.
Additionally, it is helpful to note if  $X$ and $Y$ are both symmetric trace-free tensors, then
\begin{equation} \label{teq}
 X_{IK} \widetilde{Y}^{JK} = - \widetilde{X}_{IK} Y^{JK}.
\end{equation}
Furthermore, if either one of the symmetric tensors $X$ or $Y$ is trace-free, then 
\begin{equation} \label{teqtr}
 X_{IJ} \widetilde{Y}^{IJ} = - \widetilde{X}_{IJ} Y^{IJ}.
\end{equation}

With the above definitions in mind, equation \eqref{DualCharge} can be written as
\begin{equation}
\slashed{\delta}\mathcal{\widetilde{Q}}_\xi[ \delta g, g] = {\frac{1}{16\pi G}}  \int_S d\Omega \hspace{1mm}\epsilon^{IJ} \widetilde{H}_{IJ} [\xi,  g, \delta g].
\end{equation}
We now proceed as before to substitute the metric expansions \eqref{metricexpansion} and the expression for $\xi$ given in equation \eqref{BMSgen}.

\subsection{Dual charge at $\mathcal{O}(r^{0})$} \label{Dualr0}
At leading order, we find
\begin{equation}
\slashed{\delta}\widetilde{\mathcal{I}_0} = \delta(-sD_I D_J \widetilde{B}^{IJ})+\tfrac{1}{2}s\partial_u {B}_{IJ}\delta \widetilde{B}^{IJ}.
\end{equation}
As with \eqref{BMSr0Charge}, we have an integrable piece that is in general non-zero and a non-integrable piece that vanishes if and only if $\partial_u B_{IJ}=0$, i.e.\ in the absence of Bondi news. In this case, we have an infinite set of conserved charges
\begin{equation}
\mathcal{\widetilde{Q}}_0^{\ell, m} = - {\frac{1}{16 \pi G}} \int_S d\Omega \hspace{1mm} Y_{\ell m} \hspace{1mm}D_I D_J \widetilde{B}^{IJ},
\end{equation}
which are to be viewed as the generalisation of the NUT charge \cite{godazgar2019new}; see also Ref.\ \cite{Godazgar:2019dkh}.

\subsection{Dual charge at $\mathcal{O}(r^{-1})$} \label{Dualr1}
At the next order, we find
\begin{equation}
\slashed{\delta}\widetilde{\mathcal{I}_1} = \delta(-s D_I D_J \widetilde{C}^{IJ})+\tfrac{1}{2}s(\partial_u {C}_{IJ}\delta \widetilde{B}^{IJ} -\delta C_{IJ}\partial_u \widetilde{B}^{IJ}).
\end{equation}
If we assume $T_{mm}=o(r^{-3})$, which implies equation \eqref{duCIJ}, then using equation \eqref{deltaC}, we get $\slashed{\delta}\widetilde{\mathcal{I}_1}=0$ so
\begin{equation}
\widetilde{\mathcal{I}_1} = 0.
\end{equation}
This is analogous to the $\mathcal{O}(r^{-1})$ term in Section \ref{BMSr1}, where we found that the charge is zero if strong enough fall-off conditions on the energy-momentum tensor are assumed.

\subsection{Dual charge at $\mathcal{O}(r^{-2})$} \label{Dualr2}
At the next order, we find
\begin{align} 
\slashed{\delta}\widetilde{\mathcal{I}_2} =& s D_I D_J \delta\big{(}-\widetilde{D}^{IJ}+\tfrac{1}{16}B^2 \widetilde{B}^{IJ}\big{)} \notag\\[2mm]
&+s\bigg{(}\tfrac{1}{2}(\partial_u {D}_{IJ}\delta \widetilde{B}^{IJ} -\delta D_{IJ}\partial_u \widetilde{B}^{IJ})-\tfrac{1}{16}B_{IJ}(\partial_u B^2 \delta \widetilde{B}^{IJ} -\delta B^2 \partial_u \widetilde{B}^{IJ}) \notag\\[2mm]
&+\tfrac{1}{2}\partial_u C_{IJ} \delta \widetilde{C}^{IJ}+D_I (C_{1J} \delta \widetilde{B}^{IJ})-\tfrac{1}{2}\big{(}\smallint^3+1\big{)}C_{IJ}\Box  \delta{\widetilde{B}}^{IJ}+\tfrac{1}{4}\partial_z \big{(} \Box( C_{IJ} \delta \widetilde{B}^{IJ})\big{)} \notag\\[2mm]
& +\smallint^3 C_{IJ}\delta \widetilde{B}^{IJ}-\tfrac{1}{2}D_K C_{IJ} D^K \delta \widetilde{B}^{IJ}-\smallint^3 D^K C_{JK} D_I \delta \widetilde{B}^{IJ} -\tfrac{1}{16}D_I (D_J B^2 \delta \widetilde{B}^{IJ}) \notag\\[2mm]
&-\tfrac{1}{2}D_I (B_{JK}D_L B^{KL} \delta \widetilde{B}^{IJ}) \bigg{)}.
\label{TildeI2}
\end{align}
The integrable piece has $z$ degree $N_D$, whereas from \eqref{orders} we deduce that the non-integrable piece has $z$ degree at most $N_C+1\leq N_D$. In the case $N_D>N_C+1$, we have charges
\begin{equation} \label{I2TildeCharges}
\mathcal{\widetilde{Q}}_{2,n}^{\ell, m} =- {\frac{1}{16\pi G}} \int_S d\Omega \hspace{1mm} Y_{\ell m} \hspace{1mm}D_I D_J \widetilde{D}^{IJ} \bigg{|}_{\mathcal{O}(z^n)} \hspace{5mm} \text{for} \hspace{5mm} n>N_C+1.
\end{equation}
This is analogous to the result in Section \ref{BMSr2}, where we found a set of charges \eqref{I2Charges}.  As with those charges, the existence of these conserved charges is unsurprising when  we consider the Einstein equation \eqref{duDIJ}. Thus, as in Section \ref{BMSr2}, the highest non-trivial order is $\mathcal{O}(z^{N_C+1})$, which we consider next. All terms with $z$ dependence arise from the presence of $C_{IJ}$, so we start by considering such terms. Assume that $T_{0m}=o(r^{-4})$, which implies equation \eqref{C1}. Rewriting
\begin{equation}  \label{splitforI2}
s(\partial_u {D}_{IJ}\delta \widetilde{B}^{IJ} -\delta D_{IJ}\partial_u \widetilde{B}^{IJ}) = -(\delta D_{IJ} -s \partial_u D_{IJ} ) \delta \widetilde{B}^{IJ}+\delta D_{IJ}(\delta \widetilde{B}^{IJ}-s\partial_u \widetilde{B}^{IJ})
\end{equation}
and assuming $T_{mm}=o(r^{-4})$, i.e.\ equation $\eqref{duDIJ}$, and $\eqref{deltaD}$ in the first term in \eqref{splitforI2} and \eqref{deltaB} in the second term, we get
\begin{align}
\slashed{\delta}\widetilde{\mathcal{I}_2}^{(non-int)}\big{|}_{C \hspace{1mm} \text{terms}} &=  D^{\langle I}D^{J\rangle} s\hspace{1mm} \delta \widetilde{D}_{IJ} \big{|}_{C \hspace{1mm} \text{terms}}  \notag\\[2mm]
&=\frac{1}{6}\int dz  D^{\langle I}D^{J\rangle} s \bigg{(}  s(\Box-2) \widetilde{C}_{IJ} +8D^K \widetilde{C}_{K\langle I} D_{J\rangle} s \bigg{)} +  \mathcal{O}(z^{N_C}), 
\label{I2TildeC}
\end{align}
where we have used equations \eqref{duDIJ} and \eqref{deltaD}. This is the same expression as was obtained in \eqref{I2nonintNC1}, except that the tensor field $C_{IJ}$ has been twisted. Since $\widetilde{C}^{IJ}$ is also an arbitrary symmetric, traceless tensor, we again deduce that the highest order term in $\slashed{\delta}\widetilde{\mathcal{I}_2}^{(non-int)} $ is zero if and only if $ D_{\langle I} D_{J \rangle}s = 0$, i.e.\ if $s$ is an $\ell=$ 0 or 1 spherical harmonic. Assuming this to be the case, as we can see from equation \eqref{I2TildeC}, this implies that the $C_{IJ}$ terms vanish at all orders in $z$.  Furthermore, $D_{\langle I} D_{J\rangle} s=0$ implies that \eqref{deltaB} reduces to $\delta B_{IJ}=s\partial_u B_{IJ}$. Thus, the non-integrable term in equation \eqref{TildeI2} reduces to
\begin{align}
\slashed{\delta}\widetilde{\mathcal{I}_2}^{(non-int)} =&  D^{I} \big{(}s^2(C_1^0)^J \partial_u \widetilde{B}_{IJ} -\tfrac{1}{2} s^2 B^{JK}D^L B_{KL} \partial_u \widetilde{B}_{IJ} -\tfrac{1}{16} s^2 D^J B^2 \partial_u \widetilde{B}_{IJ} \big{)} \notag\\[2mm]
&+\tfrac{1}{16}s D^{\langle I} D^{J\rangle} s \big{(}B^2 \partial_u \widetilde{B}_{IJ}+2B^{KL}\partial_u \widetilde{B}_{KL} B_{IJ}\big{)},
\end{align}
where the first line is a total derivative and so can be ignored and the second line vanishes through our choice of $s$. As was the case in Section \ref{BMSr2}, the highest order term in the non-integrable piece vanishes if and only if the non-integrable piece vanishes at all orders, with the reverse argument following from \eqref{HigherOrderMustDie} in Appendix \ref{app:Polynomials}.

With $s$ an $\ell=0$ or 1 spherical harmonic, up to total derivatives, the integrable piece becomes
\begin{align}
\delta\widetilde{\mathcal{I}_2} &= D_I D_J s \hspace{1mm} \delta\big{(} -\widetilde{D}^{IJ} +\tfrac{1}{16} B^2 \widetilde{B}^{IJ}\big{)} \notag\\[2mm]
&=D_{\langle I} D_{J\rangle} s \hspace{1mm} \delta\big{(} -\widetilde{D}^{IJ} +\tfrac{1}{16}B^2 \widetilde{B}^{IJ}\big{)} \notag\\[2mm]
&=0.
\end{align}
So as before, there is no non-trivial charge at this order.

\subsection{Dual charge at $\mathcal{O}(r^{-3})$} \label{Dualr3}
At the next order, we find that
\begin{align}
\slashed{\delta}\widetilde{\mathcal{I}_3} &= -sD_I D_J \delta\widetilde{E}^{IJ}\notag\\[2mm]
&+s\bigg{(}\tfrac{1}{2}(\partial_u {E}_{IJ}\delta \widetilde{B}^{IJ} -\delta E_{IJ}\partial_u \widetilde{B}^{IJ})+\tfrac{1}{4} D_I (3C_1^K B_{JK} \delta \widetilde{B}^{IJ} - C_1^K B^{IJ} \delta \widetilde{B}_{JK}) \notag\\[2mm]
&+D_I (C_{2J} \delta \widetilde{B}^{IJ}) -\tfrac{1}{4}\big{(}1-\partial_z\big{)} \delta \widetilde{B}^{IJ} \Box D_{IJ} -\tfrac{1}{4} \big{(}2\smallint^4+3-\partial_z\big{)} D_{IJ} \Box \delta \widetilde{B}^{IJ} +\smallint^4 \delta \widetilde{B}^{IJ} D_{IJ} \notag\\[2mm]
&-\tfrac{1}{2}\big{(}2-\partial_z\big{)} D_K D_{IJ} D^K \delta \widetilde{B}^{IJ} -\smallint^4 D_K D^{JK} D^{I} \delta \widetilde{B}_{IJ} \bigg{)} + \mathcal{O}(z^{N_C}),
 \label{I3tilde}
\end{align}
where we have used equations \eqref{duCIJ} and \eqref{deltaC} to drop all terms involving only $C_{IJ}$. The integrable piece has $z$ degree $N_E$, whereas \eqref{orders} implies that the non-integrable piece has $z$ degree at most $N_D\leq N_E$. We therefore have a set of conserved charges
\begin{equation} \label{I3TildeChargesBoring}
\mathcal{\widetilde{Q}}_{3,n}^{\ell, m} =- {\frac{1}{16\pi G}} \int_S d\Omega \hspace{1mm} Y_{\ell m} \hspace{1mm}D_I D_J \widetilde{E}^{IJ} \bigg{|}_{\mathcal{O}(z^n)} \hspace{5mm} \text{for} \hspace{5mm} n>N_D.
\end{equation}
Once again, this is unsurprising, when we consider the form of Einstein equation \eqref{duEIJ}.

Next, we consider the highest order term in $\slashed{\delta}\widetilde{\mathcal{I}_3}^{(non-int)}$ and see if it is possible to make this zero in general for a particular choice of $s(x^I)$. The highest order term is $\mathcal{O}(z^{N_D})$, but in the extreme case where $N_D=N_C+1$, it is essential that $\mathcal{O}(z^{N_C+1})$ terms also vanish. We further assume $T_{0m}=o(r^{-5})$ and $T_{mm}=o(r^{-5})$. Rewriting
\begin{equation}  \label{splitforI3}
s(\partial_u {E}_{IJ}\delta \widetilde{B}^{IJ} -\delta E_{IJ}\partial_u \widetilde{B}^{IJ}) = -(\delta E_{IJ} -s \partial_u E_{IJ} ) \delta \widetilde{B}^{IJ}+\delta E_{IJ}(\delta \widetilde{B}^{IJ}-s\partial_u \widetilde{B}^{IJ})
\end{equation}
in \eqref{I3tilde} and using equations \eqref{duEIJ} and \eqref{deltaE} in the first set of terms and \eqref{deltaB} on the second set of terms, as well as equations \eqref{C1} and \eqref{C2} gives 
\begin{align} 
\slashed{\delta}\widetilde{\mathcal{I}_3}^{(non-int)} &= D^I D^Js \hspace{1mm} \delta\widetilde{E}_{IJ} + \mathcal{O}(z^{N_C}) \notag\\[2mm]
&=  D^I D^J s \bigg{(}\widetilde{X}_{\langle IJ\rangle }\big{|}_{\mathcal{O}(z^{N_D})}+\widetilde{Y}_{\langle IJ\rangle }\big{|}_{\mathcal{O}(z^{N_C+1})}\bigg{)} + \mathcal{O}(z^{N_C}) \notag\\[2mm]
&= D^{\langle I} D^{J\rangle} s \bigg{(}\widetilde{X}_{\langle IJ\rangle }\big{|}_{\mathcal{O}(z^{N_D})}+\widetilde{Y}_{\langle IJ\rangle }\big{|}_{\mathcal{O}(z^{N_C+1})}\bigg{)} + \mathcal{O}(z^{N_C}),
\label{TildeI3Rel}
\end{align}
where 
\begin{align} 
X_{IJ} &= \tfrac{1}{3}\big{(} 2\smallint^4 +\smallint^1\big{)} s D_I D^K D_{JK} -\tfrac{1}{2}\big{(}2-\partial_z\big{)}sD_{IJ}-\tfrac{1}{2}\big{(}2\smallint^4+3-\partial_z\big{)}D_{IJ}\Box s \notag\\[2mm]
&+\tfrac{1}{3}\big{(}2\smallint^4+4\smallint^1-3\big{)}D^K D_{IK}D_J s -\big{(}2\smallint^4+1\big{)}D_I D_{JK} D^K s
\label{Xexpression}
\end{align}
and
\begin{align}  
Y_{IJ} &=\frac{1}{6} \int dz \big{(} sB^{KL}D_I D_J C_{KL} - sB_{IJ}D_K D_L C^{KL} -s B^{KL} D_K D_L C_{IJ} \notag\\[2mm]
& -2sD^KB_{JK}D^L C_{LI} + 2B_{JK}D_L C^{KL}D_I s -2B_{IJ}D_L C^{KL} D_K s  \big{)}.
\label{Yexpression}
\end{align}
Note that only the symmetric traceless part of $\widetilde{X}_{IJ}$ and $\widetilde{Y}_{IJ}$, and therefore $X_{IJ}$ and $Y_{IJ}$, need be considered.

The contributions from the $X_{IJ}$ terms and $Y_{IJ}$ terms need to vanish independently in \eqref{TildeI3Rel}, as there is no Einstein equation that relates $D_{IJ}$ and $C_{IJ}$. First, we focus on $X_{IJ}$. Use of the Ricci identity and the Schouten identity \eqref{Schouten4} allows us to rewrite $X_{IJ}$ (up to the symmetric, trace-free part) as
\begin{align}
X_{IJ} &= \tfrac{1}{3}\big{(} 2\smallint^4 +\smallint^1\big{)} s D^K D_I D_{JK} -\big{(}2\smallint^4+1\big{)}D_I D_{JK} D^K s  -\big{(}2\smallint^4+3-\partial_z\big{)}D_{JK}D^K D_I s\notag\\[2mm]
&\hspace{20mm} +\tfrac{1}{3}\big{(}2\smallint^4+4\smallint^1-3\big{)}D^K D_{JK}D_I s  -\tfrac{1}{6}\big{(}8\smallint^4+4\smallint^1+6-3\partial_z\big{)}sD_{IJ}.
\end{align}
Then using equation \eqref{gettingcoefficients}, we find that the highest order term is
\begin{align}
X_{IJ}\big{|}_{\mathcal{O}(z^{N_D})} &=\big{(}-\tfrac{1}{2} s D^K D_I D_{JK} -\tfrac{1}{2} D_I D_{JK} D^K s -\tfrac{5}{2}  D_{JK}D^K D_I s -\tfrac{5}{2} D^K D_{JK}D_I s\big{)}\big{|}_{\mathcal{O}(z^{N_D})} \notag\\[2mm]
&= D^K\big{(}-\tfrac{1}{2} s D_I D_{JK} -\tfrac{5}{2}  D_{JK} D_I s\big{)}\big{|}_{\mathcal{O}(z^{N_D})}. 
\end{align}
So in \eqref{TildeI3Rel}, the contribution from $X_{IJ}$ is
\begin{align}
\slashed{\delta}\widetilde{\mathcal{I}_3}^{(non-int)}\big{|}_{D \hspace{1mm} \text{terms}} &=  -\tfrac{1}{2}D^{\langle I} D^{J \rangle} s D^K\big{(}s D_I \widetilde{D}_{JK}+5\widetilde{D}_{JK} D_I s\big{)}\big{|}_{\mathcal{O}(z^{N_D})} +\mathcal{O}(z^{N_D-1}) \notag\\[2mm]
&= \tfrac{1}{2}D^K D^{\langle I} D^{J \rangle} s \big{(} s D_I \widetilde{D}_{JK}+5\tilde{D}_{JK} D_I s \big{)}\big{|}_{\mathcal{O}(z^{N_D})} +\mathcal{O}(z^{N_D-1}),
\end{align}
up to total derivatives. The expression $s D_I \widetilde{D}_{JK}+5\widetilde{D}_{JK} D_I s$ is an arbitrary tensor that is symmetric and trace-free on its $J, K$ indices. The contribution to $\slashed{\delta}\widetilde{\mathcal{I}_3}^{(non-int)}$ from $D_{IJ}$ terms therefore vanishes if and only if the projection of $D_K D_{\langle I}D_{J \rangle} s$ onto the symmetric trace-free part in the $JK$ indices is zero. It is shown in section 5.4 of Ref.\ \cite{godazgar2019tower} that this condition is satisfied if and only if $s$ is an $\ell=0,1$ or $2$ spherical harmonic. As in Section \ref{BMSr3}, the contribution of the $\ell=0,1$ modes vanishes in the integrable piece and so can be ignored. Henceforth, we assume that $s$ is an $\ell=2$ spherical harmonic satisfying equations \eqref{lis2} and \eqref{lis2other}.

Turning our attention to the $Y_{IJ}$ contributions in \eqref{TildeI3Rel}, the expression in \eqref{Yexpression} can be rewritten as
\begin{align}
Y_{IJ} &= -\frac{1}{3}\int dz D^K (sB_{JK} D^L C_{IL})\notag\\[2mm]
&+\frac{1}{6}s\int dz \big{(}B^{KL}D_I D_J C_{KL} -B_{IJ}D_K D_L C^{KL}-B^{KL}D_K D_L C_{IJ} + 2B_{JK}D^K D^L C_{IL} \big{)}\notag\\[2mm]
&+\frac{1}{3} \int dz \big{(}B_{JK}D_L C^{KL}D_I s -B_{IJ}D_L C^{KL}D_K s +B_{JK} D^L C_{IL} D^K s \big{)}.
\end{align}
Since $B_{IJ}$ and $C_{IJ}$ are symmetric and traceless, using Schouten identities \eqref{Schouten5} and \eqref{Schouten6}, the second and third lines have zero trace-free symmetric parts and hence can be ignored. The contribution of the $Y_{IJ}$ terms to $\slashed{\delta}\widetilde{\mathcal{I}_3}^{(non-int)}$  is therefore simply
\begin{align}
\slashed{\delta}\widetilde{\mathcal{I}_3}^{(non-int)}\bigg{|}_{C \hspace{1mm} \text{terms}} &= -\frac{1}{3}  D^{\langle I} D^{J \rangle} s \int dz\, D^K (sB_{JK}\nabla^L C_{IL})\big{|}_{\mathcal{O}(z^{N_C+1})} +\mathcal{O}(z^{N_C}) \notag\\[2mm]
&= \frac{1}{3}s D^K  D^{\langle I} D^{J \rangle} s\int dz\, B_{JK}\nabla^L C_{IL}\big{|}_{\mathcal{O}(z^{N_C+1})} +\mathcal{O}(z^{N_C}),
\end{align}
up to total derivatives. Since $B_{JK}$ is symmetric and traceless, $ D^K  D^{\langle I} D^{J \rangle} s$ is projected onto the symmetric trace-free part on its $JK$ indices, which vanishes given that $s$ is an $\ell=2$ spherical harmonic.  We conclude that for $s$ an $\ell=2$ spherical harmonic, the $\mathcal{O}(z^{N_D})$ terms in $\slashed{\delta}\widetilde{\mathcal{I}_3}^{(non-int)}$ vanish even in the extreme case $N_C+1=N_D$. 

In summary, we have a set of conserved non-trivial charges
\begin{equation} \label{I3TildeCharge}
\mathcal{\widetilde{Q}}_m =- {\frac{1}{16\pi G}} \int_S d\Omega \hspace{1mm} Y_{2m} \hspace{1mm} D_I D_J \widetilde{E}^{IJ} \bigg{|}_{\mathcal{O}(z^{N_D})} \hspace{5mm} \text{for} \hspace{2mm} m=0, \pm 1, \pm 2.
\end{equation}

We now turn our attention to the lower order terms. Again, from \eqref{HigherOrderMustDie}, if $\slashed{\delta}\widetilde{\mathcal{I}_3}^{(non-int)}$ cannot be made to vanish at a particular order, then it cannot vanish at any lower orders when the tensor fields being considered are arbitrary. We will consider $\mathcal{O}(z^{N_D-1})$ and show that the $D_{IJ}$ terms cannot be made to vanish at this order, confirming that there are no further charges at lower orders. Using \eqref{PXz}, we find that there are two independent contributions at $\mathcal{O}(z^{N_D-1})$ that must vanish independently. The first has the same form as the highest order $D_{IJ}$ terms so vanishes if and only if $s$ is an $\ell=2$ spherical harmonic. Using \eqref{gettingcoefficients}, the remaining contribution from the second term is
\begin{align}
\slashed{\delta}\widetilde{\mathcal{I}_3}^{(non-int)}\bigg{|}_{D \hspace{1mm} \text{terms}} &=  N_D\, D^{\langle I}D^{J \rangle}s \big{(} -\tfrac{3}{8} s D^K D_I \widetilde{D}_{JK}  +\tfrac{1}{8}D_I \widetilde{D}_{JK} D^K s+\tfrac{5}{4}s \widetilde{D}_{IJ}   \notag\\[2mm]
&\hspace{20mm} +\tfrac{9}{8}\widetilde{D}_{IK}D_J D^K s -\tfrac{11}{8}D^K \widetilde{D}_{IK} D_J s \big{)}\big{|}_{\mathcal{O}(z^{N_D-1})}.
\end{align}
Reorganising the terms above gives
\begin{align}
&\slashed{\delta}\widetilde{\mathcal{I}_3}^{(non-int)}\big{|}_{D \hspace{1mm} \text{terms}} =   N_D\, D^{\langle I}D^{J \rangle}s\, \big{(} \tfrac{1}{8} \big{(}s D^K D_I \widetilde{D}_{JK}  +D_I \widetilde{D}_{JK} D^K s\big{)}-\tfrac{1}{2} s D^K D_I \widetilde{D}_{JK}\notag\\[2mm]
&\hspace{30mm} +\tfrac{5}{4}s \widetilde{D}_{IJ}  -\tfrac{11}{8}\big{(}\widetilde{D}_{IK}D_J D^K s +D^K \widetilde{D}_{IK} D_J s\big{)} +\tfrac{5}{2}\widetilde{D}_{IK}D_J D^K s \big{)}\big{|}_{\mathcal{O}(z^{N_D-1})} \notag\\[2mm]
&\hspace{10mm} = N_D\, D^{\langle I}D^{J \rangle} \big{(}-\tfrac{1}{2}s D^K D_I \widetilde{D}_{JK} +\tfrac{5}{4}s \widetilde{D}_{IJ}  +\tfrac{5}{2}\widetilde{D}_{IK}D_J D^K s \big{)}\big{|}_{\mathcal{O}(z^{N_D-1})},
\end{align}
where, we have integrated by parts and used the fact that $s$ is an $\ell=2$ spherical harmonic. Applying Schouten identities, integrating by parts and applying the equations for an $\ell=2$ spherical harmonic, we obtain
\begin{equation}
\slashed{\delta}\widetilde{\mathcal{I}_3}^{(non-int)}\big{|}_{D \hspace{1mm} \text{terms}} =  N_D\, \widetilde{D}^{IJ}\big{(}-\tfrac{19}{4}s D_I D_J s +D_I sD_J s\big{)}\big{|}_{\mathcal{O}(z^{N_D-1})}+\mathcal{O}(z^{N_D-2}).
\end{equation} 
The obstruction that prevents an integrable charge existing at this order is exactly the twist of the obstruction in $\slashed{\delta}\mathcal{I}_3^{(non-int)}\big{|}_{D \hspace{1mm} \text{terms}}$ in \eqref{I3obstruction}. 

In conclusion, the set of conserved charges that can be found by considering $\slashed{\delta}\widetilde{\mathcal{I}_3}$ are given by \eqref{I3TildeChargesBoring} and \eqref{I3TildeCharge}.

\section{Relating charges to the Newman Penrose Formalism} \label{NP}
In this section, we relate the charges obtained here to quantities in the Newman-Penrose formalism \cite{newman1962approach,newman1968new, kroon1998conserved}. At $\mathcal{O}(r^{-3})$, we will see that the BMS charge and dual charge together form a generalisation of the Newman-Penrose charges for polyhomogeneous spacetimes with finite shear.

The Newman-Penrose formalism begins with a complex null frame $\{\ell,n,m,\bar{m}\}$, which we choose to be that given in \eqref{AF:frame}. Newman-Penrose scalars are then constructed by contracting tensors into null frame components.  One such set of complex scalars are the Weyl scalars, given in equation \eqref{WeylScalars}, which parameterise the ten degrees of freedom of the Weyl tensor. We reproduce these definitions here for convenience
\begin{gather}
 \Psi_0 = \ell^a m^b \ell^c m^d C_{abcd}, \quad \Psi_1 = \ell^a n^b \ell^c m^d C_{abcd}, \quad \Psi_2 = \ell^a m^b \bar{m}^c n^d C_{abcd}, \notag \\
 \Psi_3 = \ell^a n^b \bar{m}^c n^d C_{abcd}, \quad  \Psi_4 = n^a \bar{m}^b n^c \bar{m}^d C_{abcd}.
 \label{WeylScalars2}
\end{gather}
The Riemann tensor is constructed from the Weyl tensor and the Ricci tensor and the ten degrees of freedom of the Ricci tensor, which is constrained by the Einstein equation, are given by three complex and 4 real scalars. The relevant quantities here are
\begin{gather}
\Lambda = -\tfrac{1}{24} R, \quad \Phi_{11}= -\tfrac{1}{4}\ell^a n^b R_{ab} -\tfrac{1}{4}m^a \bar{m}^b R_{ab},
\end{gather}
both of which are real.  Similarly, the connection coefficients may be written in terms of twelve complex scalars.  For our purposes, we will only be interested in one such spin coefficient that parameterises the shear of the null congruence generated by the vector field $\ell$,
\begin{equation}
\sigma = -m^a m^b \nabla_b \ell_a.
\end{equation}

All such quantities can be calculated from the metric \eqref{metric}, \eqref{metricexpansion}. We assume that the energy-momentum tensor falls off as $T_{00}=o(r^{-5})$, $T_{0m}=o(r^{-4})$ and $T_{01}=o(r^{-3})$. Then one can show that the Weyl scalars fall-off as \cite{valiente1999logarithmic}
\begin{gather} 
 \Psi_0 =  \Psi_0^4[N_C]\, r^{-4}+\Psi_0^5[N_D]\, r^{-5}+\Psi_0^6[N_E]\, r^{-6}+\mathcal{O}(r^{-7}\log^{N_1} r),  \notag \\[2mm]
  \quad \Psi_1 = \Psi_1^4[N_C+1]\, r^{-4}+\mathcal{O}(r^{-5} \log ^{N_2} r),  \notag \\[2mm]
\quad \Psi_2 =\Psi_2^3[0]\, r^{-3}+\Psi_2^4[N_C+1]r^{-4}+\mathcal{O}(r^{-5}\log^{N_3} r),  \notag \\[2mm]
 \Psi_3 =\mathcal{O}(r^{-2}), \qquad \Psi_4 = \mathcal{O}(r^{-1}),
  \label{WeylScalarsFallOff}
\end{gather}
where the quantities in square brackets in each expression refer to the $z$ degree of each polynomial. The exact values of $N_1, N_2$ and $N_3$ are not important for what we are concerned with. The leading order shear term is independent of $z$
\begin{equation} \label{sigma0}
\sigma = \sigma^0[0] \frac{1}{r^2},
\end{equation}
which follows from condition \eqref{fshear}.  Furthermore,\footnote{If one assumes $T_{01}=o(r^{-4})$, then $N_4=N_C+1$.}
\begin{gather} 
\Lambda = \Lambda_4[N_4]\, r^{-4}+\mathcal{O}(r^{-5}\log^{N_5} r), \qquad
\Phi_{11} = \Phi_{11}^4\, [N_4+1] +\mathcal{O}(r^{-5}\log^{N_6} r), 
\label{RicciFalloff}
\end{gather}
where the exact values of $N_5$ and $N_6$ are unimportant.

Finally, we define the differential operators $\eth$ and $\bar{\eth}$, which act on a scalar $\eta$ of spin $n$ as follows \cite{goldberg1967spin,newman1962approach}
\begin{align}
\eth \eta &= -\tfrac{(1+i)}{2} \sin ^n \theta \big{(}\partial_\theta -\tfrac{1}{\sin \theta} \partial_\phi \big{)} \big{(} \tfrac{\eta}{\sin^n \theta}\big{)}, \ \
\bar{\eth} \eta &= -\tfrac{(1-i)}{2}\tfrac{1}{\sin^n \theta} \big{(}\partial_\theta +\tfrac{1}{\sin \theta} \partial_\phi \big{)} \big{(} \sin ^n \theta\hspace{1mm} \eta\big{)}.
\end{align}
A Weyl scalar $\Psi_n$ has spin $2-n$, while the shear $\sigma$ has spin 2. Complex conjugation changes the sign of the spin.

\subsection{Charges at $\mathcal{O}(r^{0})$}
At leading order, we obtained the BMS charges and the dual charges in Sections \ref{BMSr0} and \ref{Dualr0},
\begin{gather}
\mathcal{Q}_0^{(int)} =\frac{1}{16\pi G} \int_S d\Omega  \big{(} -2s F_0\big{)}, \quad \mathcal{\widetilde{Q}}_0^{(int)} =\frac{1}{16\pi G} \int_S d\Omega  \big{(} -sD_I D_J \widetilde{B}^{IJ}\big{)}. 
\end{gather}
Recall that the leading order charges are integrable if and only if $\partial_u B_{IJ}=0$. We define a complex quantity
\begin{equation}
\mathcal{Q}_0 = \mathcal{Q}_0^{(int)} - i \mathcal{\widetilde{Q}}_0^{(int)}. 
\end{equation} 
In terms of Newman-Penrose quantities,
\begin{equation}
\mathcal{Q}_0 = -\frac{1}{4\pi G} \int_S d\Omega  \hspace{1mm}s \big{(}\Psi_2^3 +\sigma^0 \partial_u \bar{\sigma}^0\big{)},
\end{equation} 
which is conserved if and only if $\partial_u \sigma^0=0$.  This condition is equivalent to $\partial_u B_{IJ}=0$; the integrability condition encountered in Sections \ref{BMSr0} and \ref{Dualr0}.

\subsection{Charges at $\mathcal{O}(r^{-1})$}
In sections \ref{BMSr1} and \ref{Dualr1}, assuming that $T_{mm}=o(r^{-3})$ and $T_{0m}=o(r^{-4})$, we obtained the following set of integrable charges at the next order\footnote{The dual charge here is trivially conserved and vanished in Section \ref{Dualr2} by virtue of the fact that $\delta C_{IJ}=0$.}
\begin{gather}
\mathcal{Q}_1^{(int)}(z) =\frac{1}{16\pi G} \int_S d\Omega  \hspace{1mm} s\big{(}    -2F_1-(1-\partial_z)D_I C_1^I+\tfrac{3}{16}(\Box-2)B^2 \notag\\[2mm]
\hspace{40mm}+D_IB^{IJ}D^KB_{JK}-\tfrac{1}{4}D_I B_{JK} D^I B^{JK}   \big{)}, \\[2mm]
\mathcal{\widetilde{Q}}_1^{(int)} (z) =\frac{1}{16\pi G} \int_S d\Omega   \hspace{1mm} \big{(}-sD_I D_J \widetilde{C}^{IJ}\big{)}.
\end{gather}
Note that the coefficient of each power of $z$ is an independent charge. Letting
\begin{equation}
\mathcal{Q}_1 = \mathcal{Q}_1^{(int)} - i \mathcal{\widetilde{Q}}_1^{(int)},
\end{equation}
it can be shown that
\begin{equation}
\mathcal{Q}_1 = \frac{1}{4\pi G} \int_S d\Omega  \hspace{1mm}s \bigg{(}\big{(}\smallint^2-\smallint^1\big{)} \bar{\eth}^2 \Psi_0^4 -\smallint^1 \big{(}\Phi_{11}^4 +3 \Lambda_4\big{)} \bigg{)}.
\end{equation}
The first term is trivially conserved since $T_{mm}=o(r^{-3})$ implies $\partial_u \Psi_0^4=0$. Assuming $T_{01}=o(r^{-4})$ makes the second term zero. The second term is real and gives the non-trivial conserved charges \eqref{I1Charges} in Section \ref{BMSr1} when the fall-off of the energy-momentum tensor is not too strong.

\subsection{Charges at $\mathcal{O}(r^{-2})$}
At the next order, in Sections \ref{BMSr2} and \ref{Dualr2}, we obtained the charges
\begin{gather}
\mathcal{Q}_2^{(int)}(z) =\frac{1}{16\pi G} \int_S d\Omega  \hspace{1mm} sD_I D_J\big{(} -D^{IJ}+\tfrac{1}{16}B^2 B^{IJ}\big{)}, \\[2mm]
 \mathcal{\widetilde{Q}}_2^{(int)}(z) =\frac{1}{16\pi G} \int_S d\Omega   \hspace{1mm} sD_I D_J\big{(} -\widetilde{D}^{IJ}+\tfrac{1}{16}B^2 \widetilde{B}^{IJ}\big{)}
\end{gather}
and showed that the associated non-integrable terms vanished for $s$ an $\ell=0,1$ spherical harmonic. It can be shown that
\begin{equation}
\bar{\eth}^2 \Psi_0^5  = D_I D_J \bigg{(} -\tfrac{1}{4} \big{(} \partial_z^2 -5\partial_z +6\big{)} \big{(}D^{IJ}-i \widetilde{D}^{IJ} \big{)} +\tfrac{3}{32} B^2 (B^{IJ} -i \widetilde{B}^{IJ}\big{)} \bigg{)}
\end{equation}
and 
\begin{equation}
 \left[ -\tfrac{1}{4}(\partial_z^2-5 \hspace{1mm}\partial_z +6) \right]^{-1}= -4(\smallint^3-\smallint^2)
\end{equation}
as an operator equation.  Thus,
\begin{equation}
-4\big{(}\smallint^3-\smallint^2\big{)}\bar{\eth}^2 \Psi_0^5  = D_I D_J \bigg{(} \big{(}D^{IJ}-i \widetilde{D}^{IJ} \big{)} -\tfrac{1}{16} B^2 (B^{IJ} -i \widetilde{B}^{IJ}\big{)} \bigg{)},
\end{equation}
where we have used equation \eqref{fshear}, i.e.\ that $B_{IJ}$ is $z$-independent.
Defining
\begin{equation}
\mathcal{Q}_2(z) = \mathcal{Q}_2^{(int)}(z) - i \mathcal{\widetilde{Q}}_2^{(int)}(z),
\end{equation}
it can be shown that the charges obtained in Sections \ref{BMSr2} and \ref{Dualr2} at $\mathcal{O}(z^{N_C+1})$ and lower can be written in terms of Newman-Penrose quantities as
\begin{equation}
\mathcal{Q}_2(z) =  \frac{1}{4\pi G} \int_S d\Omega   \hspace{1mm}s \big{(}\smallint^3-\smallint^2\big{)}\bar{\eth}^2 \Psi_0^5,
\end{equation} 
where each coefficient of a $z$ power in $\mathcal{Q}_2(z)$ is an independent conserved charge. Integrating by parts, the differential operators can be moved onto $s$ confirming that this is zero for $s$ an $\ell=0,$ 1 spherical harmonic, since $\bar{\eth}^2 Y_{\ell m}=0$ for $\ell=0,1$.

\subsection{Charges at $\mathcal{O}(r^{-3})$} \label{NPr3}
Finally, and most interestingly, in Sections \ref{BMSr3} and \ref{Dualr3}, we obtained the charges
\begin{gather}
\mathcal{Q}_3^{(int)} =\frac{1}{16\pi G} \int_S d\Omega   \hspace{1mm}s\big{(}-D_I D_JE^{IJ}\big{)}\big{|}_{\mathcal{O}(z^{N_D})}, \\[2mm]
 \mathcal{\widetilde{Q}}_3^{(int)} =\frac{1}{16\pi G} \int_S d\Omega  \hspace{1mm} s\big{(}-D_I D_J \widetilde{E}^{IJ}\big{)}\big{|}_{\mathcal{O}(z^{N_D})},
 \end{gather}
and showed that the associated non-integrable pieces vanished for $s$ an $\ell=0,1$ or 2 spherical harmonic. It can be shown that
\begin{equation} \label{ethethpsi6}
\bar{\eth}^2 \Psi_0^6  =  -\tfrac{1}{4} \big{(} \partial_z^2 -7\partial_z +12\big{)} D_I D_J \big{(}E^{IJ}-i \widetilde{E}^{IJ}\big{)} +\mathcal{O}(z^{N_C})
\end{equation}
and 
\begin{equation}
 \left[ -\tfrac{1}{4}(\partial_z^2-7 \hspace{1mm}\partial_z +12)\right]^{-1}=-4(\smallint^4-\smallint^3)
\end{equation}
as an operator equation. Thus,
\begin{equation}
-4\big{(}\smallint^4-\smallint^3\big{)}\bar{\eth}^2 \Psi_0^6  = D_I D_J \big{(} E^{IJ}-i \widetilde{E}^{IJ}\big{)}+\mathcal{O}(z^{N_C}).
\end{equation}
Defining
\begin{equation}
\mathcal{Q}_3 = \mathcal{Q}_3^{(int)} - i \mathcal{\widetilde{Q}}_3^{(int)},
\end{equation}
it can be shown that the charges obtained in Sections \ref{BMSr3} and \ref{Dualr3} can be written in terms of Newman-Penrose quantities as
\begin{equation}
\mathcal{Q}_3 =  \frac{1}{4\pi G} \int_S d\Omega \, s \, \big{(}\smallint^4-\smallint^3\big{)}\bar{\eth}^2 \Psi_0^6\big{|}_{\mathcal{O}(z^{N_D})}.
\end{equation}
Recalling that, furthermore, we have another set of less-interesting conserved charges \eqref{I3BoringCharge} and \eqref{I3TildeChargesBoring}, we readily deduce that the expression
\begin{equation} \label{higherE}
\frac{1}{16 \pi G} \int_S d\Omega\, s \left[ -D_ID_J\big{(}E^{IJ}-i\widetilde{E}^{IJ} \big{)}\big{|}_{\mathcal{O}(z^N)} \right]
\end{equation}
is a conserved charge for $N>N_D$ and any $s$, including, in particular, when $s$ is an $\ell =0,1$ or $2$ spherical harmonic. The $\partial_z$ terms in \eqref{ethethpsi6} evaluated at $\mathcal{O}(z^{N_D})$ carry contributions only from charges \eqref{higherE} and hence it is possible to produce a more simple expression for the charge built out of Newman-Penrose quantities given by
\begin{equation}
\mathcal{Q^{VK}}_3 =  \frac{1}{48\pi G} \int_S d\Omega \hspace{1mm}  Y_{2,m} \hspace{1mm} \bar{\eth}^2  \Psi_0^6\big{|}_{\mathcal{O}(z^{N_D})}  \hspace{5mm} \text{for} \hspace{2mm} m=0, \pm 1,\pm 2.
\end{equation}
Integrating by parts, we obtain the generalisation found in Ref.\ \cite{kroon1998conserved} for the Newman-Penrose charges of polyhomogeneous spacetimes with finite shear. For a smooth spacetime, $N_D=0$; hence the above expression reduces to the original Newman-Penrose charges \cite{newman1968new}.

\section*{Acknowledgements}

We would like to thank Hadi Godazgar, Chris Pope and Juan Valiente Kroon for discussions. M.G.\ is supported by a Royal Society University Research Fellowship. G.L.\ is supported by a Royal Society Enhancement Award. 

\appendix
\section{Polynomials in $z=\log r$} \label{app:Polynomials}
In this appendix, we collect some useful properties of polynomials in $z$.  For $\lambda\in \mathbb{R} \diagdown \{ 0 \}$ and $n\in \mathbb{N}_{\geq 0}$
\begin{align} 
\smallint^{\lambda} z^n &\equiv e^{\lambda z} \int dz  e^{-\lambda z} z^n  \notag \\[2mm]
&= e^{\lambda z} \bigg{(} -\frac{1}{\lambda} e^{-\lambda z} z^n +\frac{n}{\lambda} \int dz e^{-\lambda z}  z^{n-1}\bigg{)}  \notag \\[2mm]
&= e^{\lambda z} \bigg{(} -\frac{1}{\lambda} e^{-\lambda z} z^n-\frac{n}{\lambda^2}e^{-\lambda z} z^{n-1}\bigg{)}+\mathcal{O}(z^{n-2}) \notag \\[2mm]
&= -\frac{1}{\lambda} z^n -\frac{n}{\lambda^2}z^{n-1} +\mathcal{O}(z^{n-2}).
\label{znint}
\end{align}
Let $p(z)=p_n z^n + p_{n-1}z^{n-1} +\mathcal{O}(z^{n-2})$ be a polynomial in $z$, then using \eqref{znint}, we have by linearity of $\smallint^{\lambda}$ 
\begin{equation} \label{gettingcoefficients}
\smallint^{\lambda} p(z) = -\tfrac{1}{\lambda} p_n z^n -\big{(} \tfrac{n}{\lambda^2} p_n +\tfrac{1}{\lambda}p_{n-1}\big{)}z^{n-1}+ \mathcal{O}(z^{n-2}).
\end{equation}
Also,
\begin{equation}
\partial_z p(z) = n p_n z^{n-1}+ \mathcal{O}(z^{n-2}).
\end{equation}
In particular, note that for $c$ independent of $z$,
\begin{equation} \label{app:cint}
 \smallint^{\lambda} c = -\frac{1}{\lambda}\, c.
\end{equation}

If we apply a generic linear operator $\mathcal{O}$ formed of  $\partial_z$, $1$ and $\smallint^{\lambda}$ with $\lambda\neq0$ to any $p(z)=p_n z^n + p_{n-1}z^{n-1} $, we get a new polynomial $\tilde{p}(z)=\mathcal{O}p(z)$ of the same degree, which can be expressed in the form
 \begin{equation} \label{Opz}
 \tilde{p}(z)= A p_n z^n + \big{(}A p_{n-1} + nB p_n\big{)}z^{n-1}+ \mathcal{O}(z^{n-2})
 \end{equation}
with $A$ and $B$ $n$-independent constants depending on the choice of $\mathcal{O}$.

Let $X$ be some tensor of interest, depending on $(u, r, x^I)$ where the $r$ dependence is such that $X$ can be written as a polynomial in $z=\log r$ with coefficients depending on $(u,x^I)$, so $X=\sum_{i=0}^n X_i z^i$ where $X_i$ are tensors of the same rank as $X$ and independent of $z$. Taking angular derivatives of such an expression, for example $\Box X$, gives an expression of the form $F(D_I)[X]=\sum_{i=0}^n Y(X_i) z^i$, where at each order the same function $Y$ appears. Suppose we have an expression $P_X(z)$ involving the $\log r$ operators above, $X$ and its derivatives, where $X$ only appears linearly.  In general, we can decompose such as object as follows
\begin{equation} \label{generalX}
P_X(z)=\sum_{a}\mathcal{O}_aF_a(D_I)[X], 
\end{equation}
where $F_a(D_I)[X]=\sum_{i=0}^n Y_a(X_i) z^i$ for some $Y_a$. Then by equation \eqref{Opz}, we can write
\begin{equation}
\mathcal{O}_a F_a(D_I)[X] = A_a Y_a(X_n) z^n +\big{(}A_a Y_a(X_{n-1})+nB_a Y_a(X_n)\big{)}z^{n-1} + \mathcal{O}(z^{n-2}),
\end{equation}
which implies that equation \eqref{generalX} reduces to
\begin{align}
P_X(z)&= \sum_{a}\mathcal{O}_aF_a(D_I)[X] =  \sum_a A_aY_a(X_n)z^n \notag \\
&\hspace{20mm}+ \bigg{(}\sum_aA_aY_a(X_{n-1}) +n\sum_a B_aY_a(X_n)\bigg{)}z^{n-1} +\mathcal{O}(z^{n-2}).
 \label{PXz}
\end{align}
Now, we consider making this expression vanish at various orders for general $X$. This expression vanishes at the highest order $z^n$, if and only if
\begin{equation}
\sum_a A_aY_a(X_n) =0.
\end{equation}
Note that because we assume $X$ to be some arbitrary tensor, the above equation must hold as an operator equation and should not be viewed an an equation for $X_n$.  At the next order $z^{n-1}$, there are two terms that need to vanish independently since one depends only on $X_n$ and the other depends only on $X_{n-1}$, which are not necessarily related. So the $z^{n-1}$ coefficient vanishes if and only if
\begin{equation}
\sum_a A_aY_a(X_{n-1}) =0 \hspace{10mm} \text{and}\hspace{10mm}  \sum_a B_aY_a(X_n) =0.
\end{equation}
Since both $X_n$ and $X_{n-1}$ are arbitrary, these conditions show that if $P_X(z)$ vanishes at the highest order, checking that it vanishes at the second highest order only requires one to check that $\sum_a B_aY_a(X_n) =0$ for some arbitrary $X_n$. Furthermore,
\begin{equation}
P_X (z) \big{|}_{z^{n-1}}=0 \hspace{5mm}\Rightarrow\hspace{5mm} P_X (z) \big{|}_{z^{n}}=0.
\end{equation}
This argument can be extended to all orders, where at each order a new condition arises, but the previous conditions must still be met. We deduce for general $X$
\begin{equation}
P_X (z) \big{|}_{z^{i-1}}=0 \hspace{5mm}\Rightarrow\hspace{5mm} P_X (z) \big{|}_{z^{i}}=0
\end{equation}
for $i=1,...,n$. In particular, considering the contrapositive,
\begin{equation} \label{HigherOrderMustDie}
P_X (z) \big{|}_{z^{i}}\neq0 \hspace{5mm}\Rightarrow\hspace{5mm} P_X (z) \big{|}_{z^{i-1}} \neq0 \hspace{5mm} \forall\ 1\leq i \leq n,
\end{equation}
i.e.\ in order for the expression $P_X(z)$ to vanish at a particular order for general $X$, it needs to vanish at all higher orders for general $X$.

\section{Identities for tensors on the 2-sphere} \label{app:Schouten}
Schouten identities have been used extensively in this paper to simplify expressions. For a traceless, symmetric tensor $X_{IJ}$, the Schouten identity implies that \cite{godazgar2019subleading}
\begin{equation} \label{Schouten1}
\omega_{IJ}X_{KL}+\omega_{KL}X_{IJ}-\omega_{IL}X_{JK}-\omega_{JK}X_{IL}=0.
\end{equation}
This equation can be used as the starting point for deriving further useful identities.  In addition to those identities listed in appendix B of Ref.\ \cite{godazgar2019subleading}, in this appendix, we list a few other important examples.  For $X_{IJ}$ and $Y_{IJ}$ arbitrary symmetric, traceless tensors and $s$ some arbitrary scalar,
\begin{gather}
 {X_{\langle I}}^K Y_{J\rangle K} = 0, \label{Schouten2} \\[2mm]
D_{\langle I}D^K X_{J \rangle K} = \frac{1}{2}\Box X_{IJ}-X_{IJ}, \label{Schouten3} \\[2mm]
X_{K\langle I}D_{J \rangle} D^K s=\frac{1}{2} X_{IJ} \Box s, \label{Schouten4} \\[2mm]
X_{IJ}D_L Y^{KL} D_K s - X_{\langle I | K}D_L Y^{KL} D_{| J\rangle} s - X_{\langle I | K} D^L Y_{| I \rangle L} D^K s =0, \label{Schouten5} \\[2mm]
Y_{IJ}D_K D_L X^{KL} +Y^{KL}D_K D_L X_{IJ} -Y^{KL}D_{\langle I}D_{J\rangle} X_{KL} -2Y_{\langle I | K}D^K D^L X_{| J \rangle L}=0. \label{Schouten6}
\end{gather}

We briefly explain in turn how the above identities are obtained from equation \eqref{Schouten1}.  Contracting \eqref{Schouten1} with symmetric traceless tensor $Y^{KL}$ gives \eqref{Schouten2}.  Next, we apply the derivative operator $D^L D^K$ on equation \eqref{Schouten1}.  This gives
\begin{equation}
\omega_{IJ}D^K D^L X_{KL} +\Box X_{IJ} -D_I D^K X_{JK} - D^K D_J X_{IK}=0.
\end{equation}
Making use of the Ricci identity and the form of the Riemann tensor 
\begin{equation}
 {R^I}_{JKL} = \delta^I_K \omega_{JL}-\delta^I_L\omega_{JK}
\end{equation}
for the standard 2-sphere metric $\omega_{IJ}$, we get
\begin{equation}
\omega_{IJ}D^K D^L X_{KL} +\Box X_{IJ} -D_I D^K X_{JK} - D_J D^K X_{IK}-2X_{IJ}=0,
\end{equation}
which is equivalent to \eqref{Schouten3}.  This is an important identity that is used frequently in this paper.  Similarly, contracting \eqref{Schouten1} with $D_K D_L s$ gives \eqref{Schouten4}.

Contracting equation \eqref{Schouten1} with $D_MY^{KM} D^L s$ and taking the symmetric trace-free part of the resulting equation gives equation \eqref{Schouten5}.

Finally, we apply $Y^{KM} D^L D_M$ as an operator on equation \eqref{Schouten1} to obtain
\begin{equation} \label{SchoutenA}
\omega_{IJ} Y^{KM}D^L D_M X_{KL} + Y^{KL}D_K D_L X_{IJ} - Y^{KL} D_{I}D_{L} X_{JK} - Y_{JK} D^L D^K X_{IL}=0.
\end{equation}
Now, relabelling $I\rightarrow M$ in equation \eqref{Schouten1} and acting with $Y^{ML} D_I D^K$ gives
\begin{equation} \label{SchoutenBB}
Y_{JL}D_I D_K X^{KL}+Y^{KL}D_I D_L X_{JK}-Y^{KL}D_I D_J X_{KL}=0.
\end{equation}
Using equation \eqref{SchoutenBB} to substitute for the $Y^{KL}D_I D_L X_{JK}$ term in equation \eqref{SchoutenA} gives
\begin{align} 
\omega_{IJ} Y^{KM}D^L D_M X_{KL} &+ Y^{KL}D_K D_L X_{IJ}+Y_{JL}D_I D_K X^{KL}\notag\\[2mm]
&\hspace{20mm}-Y^{KL}D_I D_J X_{KL}- Y_{JK} D^L D^K X_{IL}=0.
\label{SchoutenB}
\end{align}
Next, replacing $X\rightarrow Y$ in equation \eqref{Schouten1} and acting with $D^L D_M X^{KM}$ gives
\begin{eqnarray} \label{SchoutenC}
\omega_{IJ}Y^{KM}D_M D^L X_{KL}+Y_{IJ}D_K D_L X^{KL} - Y_{JL}D_I D_K X^{KL} - Y_{IK} D^K D^L X_{JL}=0,
\end{eqnarray}
which we use to replace the $Y_{JL}D_I D_K X^{KL}$ term in equation \eqref{SchoutenB}, resulting in
\begin{eqnarray}
&\omega_{IJ} Y^{KM}D^L D_M X_{KL} + Y^{KL}D_K D_L X_{IJ}+\omega_{IJ}Y^{KM}D_M D^L X_{KL}+Y_{IJ}D_K D_L X^{KL} \notag\\[2mm]
&\hspace{20mm} - Y_{IK} D^K D^L X_{JL}-Y^{KL}D_I D_J X_{KL}- Y_{JK} D^L D^K X_{IL}=0.
 \label{SchoutenD}
\end{eqnarray}
We use the Ricci identity to exchange the $D_K$ and $D_L$ derivatives in the last term.  This results in an additional term of the form $X_{IK}{Y_{J}}^K$.  Now, taking the symmetric trace-free part of this equation and using equation \eqref{Schouten2} yields identity \eqref{Schouten6}.

\section{$\ell=0$, $\ell=1$ and $\ell=2$ spherical harmonics} \label{app:sh}
In this appendix, we list useful properties of $\ell \leq 2$ spherical harmonics. This appendix has a large overlap with appendix C of Ref.\ \cite{godazgar2019subleading}.  However, given the importance of these results in this paper, for completeness, we reproduce the relevant equations here.  A regular function $\psi(x^I)$ on the sphere can be written in terms of an expansion
\begin{equation} \label{shexpansion}
\psi(x^I) = \sum_{\ell=0}^\infty \sum_{m=-\ell}^l \psi^{\ell m}\, Y_{\ell m}(x^I)
\end{equation}
with $\psi^{\ell m}$ constants on the sphere and the spherical harmonics $Y_{\ell m}(x^I)$ with $\ell \geq0$ and $|m|\leq \ell$ obeying 
\begin{equation}
 \Box Y_{\ell m} = -\ell(\ell+1) Y_{\ell m}.
\end{equation}

Consider the equation
\begin{equation} \label{psil2}
D_{\langle I} D_{J \rangle} \psi = 0.
\end{equation}
Let $T_{IJ}=D_{\langle I} D_{J \rangle} \psi$. Integrating by parts and using the Ricci identity, it can be shown
\begin{equation} \label{Tsq}
\int_S d\Omega\hspace{1mm}  |T_{IJ}|^2 = \frac{1}{2} \int_S d\Omega \hspace{1mm}   \psi\Box (\Box+2)\psi.
\end{equation}
If $\psi(x^I)$ is regular, we can assume the expansion \eqref{shexpansion}. Plugging this into \eqref{Tsq} and using the orthogonality relations for spherical harmonics $\int_S d\Omega\hspace{1mm}  Y_{\ell m} Y_{\ell^{\prime}m^{\prime}} = \delta_{\ell \ell^{\prime}}\delta_{m m^{\prime}}$, we find
\begin{align}
\int_S d\Omega\hspace{1mm}  |T_{IJ}|^2 &= \frac{1}{2} \int_S d\Omega \hspace{1mm}  \sum_{\ell=0}^\infty \sum_{\ell^{\prime}=0}^\infty\sum_{m=-\ell}^\ell   \sum_{m^{\prime}=-\ell^{\prime}}^{\ell^{\prime}} (\ell-1)\ell(\ell+1)(\ell+2) \psi^{\ell m} \psi^{\ell^{\prime} m^{\prime}} Y_{\ell m} Y_{\ell^{\prime}m^{\prime}} \notag\\[2mm]
&=\frac{1}{2}   \sum_{\ell=0}^\infty \sum_{m=-\ell}^\ell (\ell-1)\ell(\ell+1)(\ell+2)|\psi^{\ell m}|^2.
\label{TsqExpansion}
\end{align} 
Notice that each term in the summation on the RHS is non-negative.  Therefore, for the RHS to vanish, all terms must individually vanish, which implies that the RHS vanishes if and only if $\psi^{\ell m}=0$ for all $\ell>1$, i.e.\ $\psi(x^I)$ is a linear combination of $\ell=0$ and $\ell=1$ modes.  We conclude, then, that equation \eqref{psil2} holds if and only if $\psi(x^I)$ is a linear combination of $\ell=0$ and $\ell=1$ modes.

Consider now equation \eqref{I3sequation}, which is equivalent to
\begin{equation}
 T_{IJK} = 0, \qquad T_{IJK}= 2D_K D_{\langle I} D_{J \rangle} \psi -\omega_{K\langle I} D_{J\rangle} \Box \psi -2\omega_{K\langle I} D_{J\rangle}  \psi. 
\end{equation}
For a function $\psi(x^I)$ that is regular on the sphere, integration by parts can be used to show that
\begin{equation}
\int_S d\Omega\hspace{1mm}  |T_{IJK}|^2 = - \int_S d\Omega \hspace{1mm}   \psi\Box (\Box+2)(\Box+6)\psi.
\end{equation}
Inserting expansion \eqref{shexpansion} into the above equation yields
\begin{equation}
\int_S d\Omega\hspace{1mm}  |T_{IJK}|^2 =  \sum_{\ell=0}^{\infty}\sum_{\ell=-m}^m    (\ell-2)(\ell-1)\ell(\ell+1)(\ell+2)(\ell+3)  |\psi^{\ell m}|^2.
\end{equation}
Using the same argument as above, we deduce that $T_{IJK}=0$ if and only if $\psi(x^I)$ is a linear combination of $\ell=0,1$ and 2 spherical harmonics.

\bibliographystyle{utphys}
\bibliography{RefsPaper1c.bib}

\end{document}